\documentclass[final,3p,times]{elsarticle}

\usepackage{amsmath, amssymb}
\usepackage{color}

\usepackage[colorlinks=true,citecolor=blue,urlcolor=blue,linkcolor=blue,hyperfigures=true]{hyperref}

\newcommand{\idmatrix}{{\mathbf{1}}}

\newcommand{\kvec}[1]{{\mathbf{#1}}}
\newcommand{\cvec}[1]{{\mathrm{#1}}}

\newcommand{\pr}{^\prime}
\newcommand{\svek}{\mathbf}
\newcommand{\out}[1]{}

\usepackage{amssymb}

\usepackage{listings}

\newcounter{bla}

\usepackage{lipsum} 

\journal{Computer Physics Communications}

\begin{document}

\begin{frontmatter}



\title{The AbinitioD$\Gamma$A Project v1.0:
Non-local correlations beyond and susceptibilities within dynamical mean-field theory}


\author[a,b]{Anna Galler\corref{author}}
\author[a,c]{Patrik Thunstr\"om}
\author[a]{Josef Kaufmann}
\author[a]{Matthias Pickem\corref{author}}
\author[a]{Jan M. Tomczak}
\author[a]{Karsten Held}

\cortext[author] {Corresponding authors.\\\textit{E-mail addresses:} galler.anna@gmail.com, matthias.pickem@gmail.com}
\address[a]{Institute of Solid State Physics, TU Wien, 1040 Vienna, Austria}
\address[b]{Centre de Physique Th\'eorique, Ecole Polytechnique, 91128 Palaiseau, France}
\address[c]{Department of Physics and Astronomy, Materials Theory, Uppsala University, 75120 Uppsala, Sweden}

\begin{abstract}
 The {\em ab initio} extension of the dynamical vertex approximation (D$\Gamma$A) method allows for realistic materials calculations that include non-local correlations beyond $GW$ and dynamical mean-field theory. Here, we discuss the AbinitioD$\Gamma$A algorithm, its implementation and usage in detail, and make the program package available to the scientific community.
\end{abstract}


\end{frontmatter}



\noindent
{\bf PROGRAM SUMMARY}

\begin{small}
\noindent
{\em Program Title:} AbinitioD$\Gamma$A                                          \\
{\em Licensing provisions:} GNU General Public License (GPLv3)                                   \\
{\em Operating system:} Linux, Unix, macOS \\
{\em Programming language:} \verb=Fortran95= and \verb=Python=                                   \\
{\em Required dependencies:} \verb=MPI=, \verb=LAPACK=, \verb=BLAS=, \verb=HDF5= $(\geq 1.8.12)$, \verb=Python= $(\geq 2.7)$, \verb=h5py= $(\geq 2.5.0)$, \verb=numpy= $(\geq 1.9.1)$ \\
{\em Optional dependencies:} \verb=pip=, \verb=matplotlib= $(\geq 1.5.1)$, \verb=scipy= $(\geq 0.14.0)$\\
{\em Supplementary material:}        Test case files and step-by-step instructions

{\em Nature of problem:}\\
Realistic materials calculations including non-local correlations beyond dynamical mean-field theory (DMFT) as well as non-local interactions. Solving the Bethe-Salpeter equation for multiple orbitals. Determining momentum-resolved susceptibilities in DMFT.

{\em Solution method:}\\
{\em Ab initio} dynamical vertex approximation: starting from the local two-particle vertex and constructing from it the local DMFT correlations, the $GW$ diagrams, and further non-local correlations, e.g., spin fluctuations. Efficient solution of the Bethe-Salpeter equation, avoiding divergencies in the irreducible vertex in the particle-hole channel by reformulating the problem in terms of the full vertex. Parallelization with respect to the bosonic frequency and transferred momentum.

{\em Additional comments including Restrictions and Unusual features:}\\
As input, a Hamiltonian derived, e.g., from density functional theory and a DMFT solution thereof is needed including a local two-particle vertex calculated at DMFT self-consistency.
Hitherto the AbinitioD$\Gamma$A program package is restricted to SU(2) symmetric problems. A so-called $\lambda$ correction or self-consistency is not yet implemented in the AbinitioD$\Gamma$A code. Susceptibilities are so far only calculated within DMFT, not the dynamical vertex approximation.
   \\

\end{small}

\section{Introduction}
\label{intro}
\noindent
Dynamical mean-field theory (DMFT)
 \cite{Metzner1989,Georges1992a} takes into account a major part of the electronic correlations, namely the local ones. It has been very successfully applied
to models of strongly correlated electron system, see \cite{Georges1996} for an early review and \cite{DMFT25} for a series of lecture notes on the occasion of 25 years of DMFT. Its merger with density functional theory (DFT) \cite{Anisimov1997,Lichtenstein1998} and $GW$~\cite{Biermann2003,Sun02}
even allows for the realistic calculation of materials including strong electronic correlations, see \cite{Kotliar2006,Held2007} and \cite{Tomczak2017review} for reviews.

On the other hand, non-local correlations are at the heart of many fascinating phenomena of many-body physics. In the aforementioned $GW$+DMFT approach
the screening of the bare interaction $V$ to a screened $W$ gives rise
to non-local correlations in the self-energy. But there are important further effects of non-local correlations, e.g., spin fluctuations.
Hence, extensions of DMFT that include the local DMFT correlations
and additional non-local correlations are at the scientific frontier.

One main route to this end are
cluster extensions of DMFT which consider a cluster of sites in a DMFT Weiss field. Two methods, the dynamical cluster approximation (DCA) \cite{Hettler1998} and the cellular DMFT \cite{Lichtenstein2000,Kotliar2001}, have been developed, see \cite{Maier2005} for a review. Due to numerical limitations these approaches are restricted to
short-range correlations and essentially a single interacting band.
 Realistic calculations are hardly possible or restricted to extremely small clusters \cite{Biermann2005,Lee12}.

Diagrammatic extensions of the DMFT on the other hand use a local two-particle vertex as a starting point and construct from it the local DMFT correlations as well as non-local correlations. These diagrammatic extensions are more suitable
to deal with long-range correlations and realistic multi-orbital calculations.
This more recent development started with the
dynamical vertex approximation (D$\Gamma$A) \cite{Toschi2007,Kusunose2006},
subsequently followed by various other approaches such as the dual fermion (DF) approach \cite{Rubtsov2008}, the one-particle irreducible (1PI) approach \cite{Rohringer2013}, the dynamical mean-field theory to functional renormalization group (DMF$^2$RG) approach \cite{Taranto2014}, the triply-irreducible local expansion (TRILEX) \cite{Ayral2015} and the non-local expansion scheme \cite{Li2015}.
These diagrammatic extensions of DMFT have been first applied to model systems, among others to
calculate (quantum) critical exponents \cite{Rohringer2011,Hirschmeier2015,Antipov2014,Schaefer2016}, see \cite{RMPvertex} for a review.

Most recently, these diagrammatic approaches have been extended to realistic multi-orbital {\em ab initio} D$\Gamma$A calculations \cite{Galler2017,JPSJ-XXX}. Using the local three-frequency and four-orbital vertex and on top of this the non-local bare interaction as a starting point, this approach not only includes the DMFT and $GW$ Feynman diagrams but also many further non-local correlations.
It is the aim of this paper to make the developed AbinitioD$\Gamma$A program package available to the general scientific community.

The paper is organized as follows:
In Sec.~\ref{Sec:equations}, we recapitulate the
AbinitioD$\Gamma$A formalism of Reference~\cite{Galler2017}
in the context of our implementation.
Subsequently we explain the program from a user's point of view. Specifically, Sec.~\ref{Sec:installation} shows the installation procedure,
Sec.~\ref{Sec:Setup} makes the reader familiar with
the necessary steps for a simple calculation, which is
illustrated in Sec.~\ref{Sec:example} by an example case: SrVO$_3$.
Further information for advanced users can be found in
Sec.~\ref{sec:adv-user-opt}, where the parameters for the example case
and the structure of the output file are discussed.
Following this description for users we switch to a more technical description in Sec.~\ref{sec:programalgo}
where we provide more details about the internal program flow and the implementation.
We provide \ref{app:configs} where further user options are discussed; and \ref{app:results} where results for a quick SrVO$_3$ test calculation with a small frequency box are presented.
Finally, Section \ref{Sec:Conclusion} provides a summary and recapitulates the approximations used when doing AbinitioD$\Gamma$A calculations.

\section{Implemented AbinitioD$\Gamma$A equations}
\label{equations}
\label{Sec:equations}
\subsection{AbinitioD$\Gamma$A self-energy}
\label{Sec:selfenergy}
\noindent
The main quantity, which is computed by the AbinitioD$\Gamma$A algorithm, is the non-local (momentum-dependent) and dynamical (frequency-dependent) 
self-energy $\Sigma_{{\mathrm{D}}\Gamma{\mathrm{A}}}$.
We compute $\Sigma_{{\mathrm{D}}\Gamma{\mathrm{A}}}$ in the ladder approximation of D$\Gamma$A ~\cite{Katanin2009,Rohringer2016} which incorporates non-local ladder diagrams
in both the particle-hole ($ph$) and transverse particle-hole ($\overline{ph}$) channel,
starting from a local irreducible vertex in these channels.
This way, among others, spin fluctuations are included, but one neglects the particle-particle ($pp$) channel which is, e.g., important for superconducting fluctuations.

In AbinitioD$\Gamma$A the local irreducible vertex in the $ph$ and $\overline{ph}$ channel is supplemented by the bare non-local Coulomb interaction $V^{\kvec{q}}$.
This generates additional diagrams and screening effects.
If one included only the $ph$ channel and the non-local Coulomb interaction, one would reproduce the $GW$ approximation \cite{Hedin1965}.

\begin{figure}[]
\centering
\includegraphics[width=12cm]{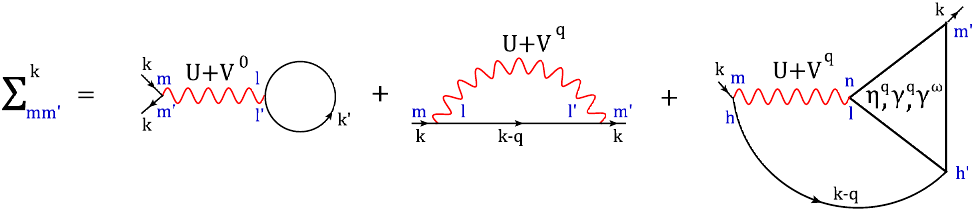}
\caption{Feynman-diagrammatic representation of the AbinitioD$\Gamma$A equation of motion [Eq.~\eqref{eq:eom_final_imp}]: The self-energy is obtained from the three-leg vertices $\eta^{\cvec{q}}$, $\gamma^{\cvec{q}}$ and $\gamma^{\omega}$, which in turn are obtained from the full vertex $F$ (see Fig.~\ref{fig:three-leg} below) that is determined via the Bethe-Salpeter equation. In addition, the Hartree and Fock contributions of Eq.~\eqref{eq:eom_hartree} need to be added. Beside the momenta $\cvec{qkk}\pr$, orbital indices $lmm\pr l\pr$ are shown in blue.}
\label{fig:eom}
\end{figure}

Let us here start the discussion of the AbinitioD$\Gamma$A equations with the final quantity calculated in the code: the self-energy.
Using the compound indices $\cvec{k}=(\kvec{k},\nu)$ and $\cvec{q}=(\kvec{q},\omega)$
for momenta $\kvec{k}$, $\kvec{q}$ and Matsubara frequencies $\nu$, $\omega$,
the self-energy of the AbinitioD$\Gamma$A is obtained from the Schwinger-Dyson equation as \cite{Galler2017}
\begin{align}
\label{eq:eom_final_imp}
\Sigma^{\cvec{k}}_{\substack{{\mathrm{D}}\Gamma{\mathrm{A}}\\ mm\pr}} = & \Sigma^{\nu}_{\substack{{\mathrm{DMFT}}\\mm\pr}}+\Sigma^{\kvec{k}}_{\substack{{\mathrm{HF}}\\mm\pr}} -\beta^{-1} \sum_{lnhh\pr \cvec{q}} \! \Big( U_{mlhn} + V^{\kvec{q}}_{mlhn} - \frac{1}{2}\tilde{U}_{mlhn}\Big)\eta^{\cvec{q}\nu}_{d,nlh\pr m\pr}G^{\cvec{k}-\cvec{q}}_{hh\pr}\\ \nonumber & +\beta^{-1} \sum_{lnhh\pr\cvec{q}} \! \frac{3}{2}\tilde{U}_{mlhn}\eta^{\cvec{q}\nu}_{m,nlh\pr m\pr}G^{\cvec{k}-\cvec{q}}_{hh\pr} -\beta^{-1} \sum_{lnhh\pr\cvec{q}} \! \Big( V^{\kvec{q}}_{mlhn}\gamma^{\omega\nu}_{d,nlh\pr m\pr}- U_{mlhn}\gamma^{\cvec{q}\nu}_{d,nlh\pr m\pr}\Big)G^{\cvec{k}-\cvec{q}}_{hh\pr}.
\end{align}
This expression for the AbinitioD$\Gamma$A self-energy is depicted diagrammatically in Fig.~\ref{fig:eom}.
In the following we will introduce all quantities necessary for the evaluation of $\Sigma_{{\mathrm{D}}\Gamma{\mathrm{A}}}$
in Eq.~\eqref{eq:eom_final_imp} and discuss how they are calculated.
In particular, the non-local three-leg vertices $\gamma^{\cvec{q}\nu}$ and $\eta^{\cvec{q}\nu}$
are obtained in D$\Gamma$A from the local irreducible vertex through the Bethe-Salpeter ladder.
A brief description is given below, while a detailed derivation can be found in Ref. \cite{Galler2017}.

\paragraph*{One-particle Green's function} The one-particle Green's function $G^{\cvec{k}-\cvec{q}}$ appearing in the last three terms in Eq.~\eqref{eq:eom_final_imp} is the lattice Green's function, given by
\begin{equation}
\label{eq:g_nonloc}
G^{\cvec{k}}_{hh\pr}=\Big[i\nu+\mu-H_W^{\kvec{k}}-\Sigma^{\nu}-\Sigma_{DC}\Big]^{-1}_{hh\pr}.
\end{equation}
Here, $H_W^{\kvec{k}}$ is the material-dependent Hamiltonian obtained, e.g., from a DFT computation and a consecutive Wannier projection \cite{Kuneifmmodecheckselsevsfi2010a}, $\Sigma^{\nu}$ is the dynamical but local DMFT self-energy, $\Sigma_{DC}$ the double-counting correction, and $\mu$ the chemical potential of the DMFT calculation. Here, and in the following, Roman subscripts denote orbital indices, $\nu=(2n+1)\pi/\beta$ ($\omega=2n\pi/\beta$) are fermionic (bosonic) Matsubara frequencies for an inverse temperature $\beta=1/(k_B T)$.

From the products of two interacting one-particle Green's functions the unconnected (bare bubble) susceptibilities $\chi_0^{\omega}$, $\chi_0^{\cvec{q}}$ and $\chi_0^{nl,\cvec{q}}$ are obtained. In order of increasing non-local character, these are defined as:
\begin{align}
\chi^{\omega\nu\nu}_{0,lmm\pr l\pr} & = -\beta G^{\nu}_{ll\pr} G^{\nu-\omega}_{m\pr m}, \label{eq:bubble_loc} \\ 
\chi^{\cvec{q}\nu\nu}_{0,lmm\pr l\pr} & = -\beta \sum_{\kvec{k}}G^{\cvec{k}}_{ll\pr}G^{\cvec{k}-\cvec{q}}_{m\pr m}, \label{eq:q_bubble}\\
\chi^{\mathrm{nl},\cvec{q}\nu\nu}_{0,lmm\pr l\pr} & = \chi^{\cvec{q}\nu\nu}_{0,lmm\pr l\pr}-\chi^{\omega\nu\nu}_{0,lmm\pr l\pr}. \label{eq:bubble_nl}
\end{align}
Here, $\chi^{\omega\nu\nu}_{0}$ is a purely local bubble-term obtained by the product of two local one-particle DMFT Green's functions $G^\nu_{hh\pr}=\sum_{\mathbf k}G^{\cvec{k}}_{hh\pr}$; $\chi^{\cvec{q}\nu\nu}_{0}$ instead is the q-dependent product of two non-local DMFT Green's functions defined in Eq.~\eqref{eq:g_nonloc}. By subtracting Eq.~\eqref{eq:bubble_loc} from Eq.~\eqref{eq:q_bubble} one finally obtains the purely non-local $\chi^{\mathrm{nl},\cvec{q}\nu\nu}_{0}$.

\paragraph*{Local and non-local Coulomb interaction} In the Wannier basis, local and non-local Coulomb interaction are four-index objects, denoted as $U_{lm\pr ml\pr}$ and $V^{\kvec{q}}_{lm\pr ml\pr}$, respectively.
The current implementation allows for an arbitrary orbital-dependence of the Coulomb interaction without any restriction in performance.

\begin{figure}[]
\centering
\includegraphics[width=7.5cm]{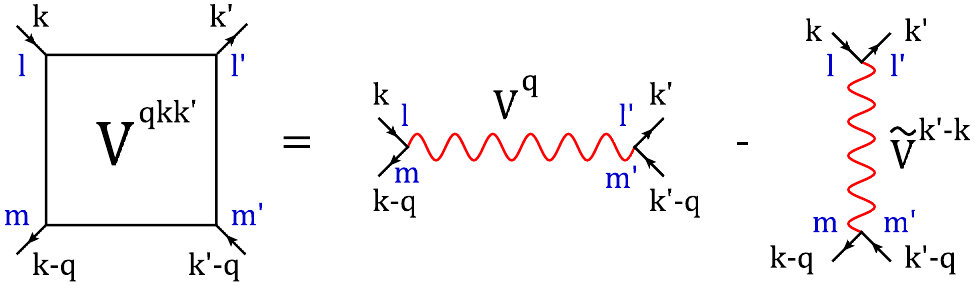}
\caption{The non-local Coulomb interaction $V^{\kvec{q}\kvec{k}\kvec{k'}}$ consists of two terms: $V^{\kvec{q}}$ and $V^{\kvec{k'}-\kvec{k}}$ (the latter is neglected in the AbinitioD$\Gamma$A approach). The local Coulomb interaction $U$ has the same structure, but without any momentum dependence so that both terms can be taken into account without effort.}
\label{fig:v-nonloc}
\end{figure}

The $\tilde{U}$ in Eq.~\eqref{eq:eom_final_imp} is the local Coulomb interaction in the transverse particle-hole ($\overline{ph}$) channel and is related to $U$ through $\tilde{U}_{lm\pr ml\pr}=U_{lm\pr l\pr m}$.
The two channels are visualized in Fig.~\ref{fig:v-nonloc} for the non-local Coulomb interaction. However, the non-local component
in the $\overline{ph}$-channel, $\tilde{V}^{\kvec{k'}-\kvec{k}}_{lm\pr l\pr m}$, is neglected in the AbinitioD$\Gamma$A formalism (it would result in a ${\kvec{k'}}$- and $\kvec{k}$-dependence in the Bethe-Salpeter ladder equations, for details see Ref.~\cite{Galler2017,JPSJ-XXX}). This approximation is common practice: Indeed the {\it GW}~\cite{Hedin1965} and {\it GW}+DMFT~\cite{Biermann2003,Sun02} approaches neglect both local and non-local interactions in the transverse channel.

\paragraph*{DMFT and Hartree-Fock contribution}
The first contribution to $\Sigma^{\cvec{k}}_{{\mathrm{D}}\Gamma{\mathrm{A}}}$ in Eq.~\eqref{eq:eom_final_imp} is the DMFT self-energy $\Sigma^{\nu}_{{\mathrm{DMFT}}}$ that contains
all diagrams that can be build from the local Green's function $G^\nu=\sum_\kvec{k}G^{\cvec{k}}$ and the local interaction $U$. This contribution is the leading term to $\Sigma^{\cvec{k}}_{{\mathrm{D}}\Gamma{\mathrm{A}}}$ at large frequencies $\nu$.
$\Sigma^{\nu}_{{\mathrm{DMFT}}}$ includes the local and static Hartree-Fock term originating from the local Coulomb interaction $U$. However, AbinitioD$\Gamma$A also contains the non-local Hartree-Fock term arising from the non-local Coulomb interaction $V^{\kvec{q}}$. Indeed, the second term in Eq.~\eqref{eq:eom_final_imp}, $\Sigma_{\mathrm{HF~}}^{\kvec{k}}$, is the non-local, static Hartree-Fock contribution and reads
\begin{equation}
\label{eq:eom_hartree}
\Sigma_{\substack{\mathrm{HF~}\\mm\pr}}^{\kvec{k}} = 2\sum_{ll\pr \svek{k\pr}}V_{mlm\pr l\pr}^{\svek{q}=\svek{0}}n^{\svek{k\pr}}_{l\pr l} - \sum_{ll\pr\svek{q}}V^{\svek{q}}_{mll\pr m\pr}n^{\svek{k}-\svek{q}}_{l\pr l} ,
\end{equation}
where $n^{\svek{k}}=1/\beta\sum_\nu{G^{\cvec{k}}}$ are the k-dependent occupancies.

\begin{figure}[]
\centering
\includegraphics[width=6.5cm]{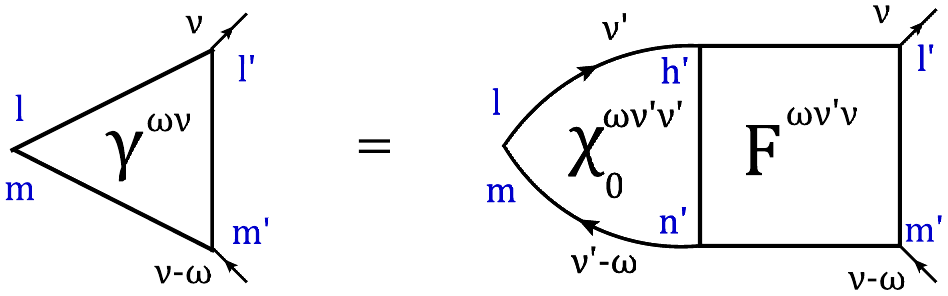}
\caption{The three-leg vertices are obtained from $\chi_0F$ through a sum over the left fermionic frequency $\nu'$. Here, exemplary, the relation for the purely local three-leg vertex $\gamma^{\omega\nu}$ in Eq.~\eqref{eq:gamma_loc_imp} is shown.}
\label{fig:three-leg}
\end{figure}

\paragraph*{Three-leg vertices} The quantities $\gamma^{\omega\nu}_r$, $\gamma^{\cvec{q}\nu}_r$ and $\eta^{\cvec{q}\nu}_r$ in Eq.~\eqref{eq:eom_final_imp} (with the index $r$ referring to the (d)ensity or (m)agnetic channel) are three-leg vertices. Such three-leg vertices have been used in D$\Gamma$A and dual fermion from the beginning, see, e.g., Refs.~\cite{Katanin2009,Hafermann2014a}; the TRILEX approach \cite{Ayral2015}, which does not solve the non-local Bethe-Salpeter equation, is formulated entirely in terms of these three-leg vertices. They can be obtained from the full, four-leg vertex function $F$ through a sum over the left fermionic frequency $\nu\pr$. A diagrammatic representation of these three-leg vertices is shown in Fig.~\ref{fig:three-leg}. With increasing order of non-locality the necessary three-leg vertices read
\begin{align}
\gamma^{\omega\nu}_{r,lmm\pr l\pr} & = \sum_{n\pr h\pr \nu'} \chi^{\omega\nu'\nu'}_{0,lmn\pr h\pr} F^{\omega\nu'\nu}_{r,h\pr n\pr m\pr l\pr}, \label{eq:gamma_loc_imp}\\
\gamma^{\cvec{q}\nu}_{r,lmm\pr l\pr} & = \sum_{n\pr h\pr\nu'} \chi^{\mathrm{nl},\cvec{q}\nu'\nu'}_{0,lmn\pr h\pr} F^{\omega\nu'\nu}_{r,h\pr n\pr m\pr l\pr},\label{eq:gamma_q_imp}\\
\eta^{\cvec{q}\nu}_{r,lmm\pr l\pr} & = \sum_{n\pr h\pr\nu'} \chi^{\cvec{q}\nu'\nu'}_{0,lmn\pr h\pr} F^{\cvec{q}\nu'\nu}_{r,h\pr n\pr m\pr l\pr} - \gamma^{\omega\nu}_{r,lmm\pr l\pr}.
\label{eq:eta_imp}
\end{align}
Thus, $\gamma^{\omega\nu}$ is a completely local three-leg vertex which can directly be extracted from the impurity solver \cite{PhysRevB.89.235128,kaufmann_vertex_asymp_2017}. If the DMFT impurity solver does not explicitly provide $\gamma^{\omega\nu}$, the latter is computed within the AbinitioD$\Gamma$A program according to Eq.~\eqref{eq:gamma_loc_imp}.

\begin{figure}[]
\centering
\includegraphics[width=6.5cm]{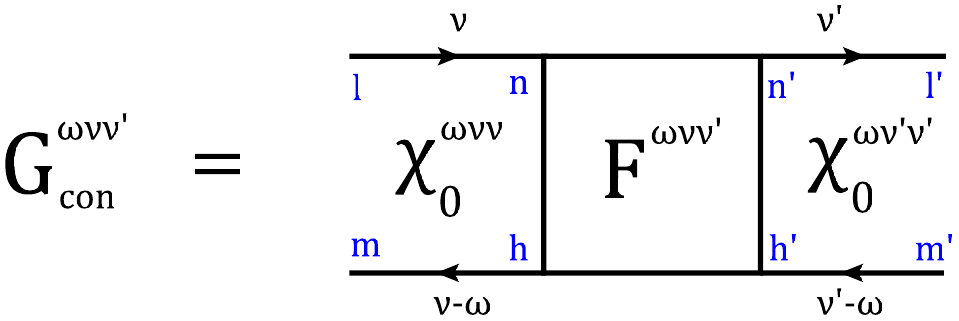}
\caption{The local vertex function $F^{\omega\nu\nu'}$ can be obtained from the connected part of the two-particle impurity Green's function $G^{\omega\nu\nu'}_{\mathrm{con}}$ through an 'amputation' of the left and the right legs [see Eq.~\eqref{eq:ffromgcon}]. }
\label{fig:g_conn}
\end{figure}

The most complex ingredients of Eqs. \eqref{eq:gamma_loc_imp}-\eqref{eq:eta_imp} are the local four-leg
vertex function $F^{\omega\nu'\nu}_{r}$ of the DMFT impurity and its non-local counterpart
$F^{q\nu'\nu}_{r}$. The local four-leg vertex $F^{\omega\nu'\nu}_{r}$ is obtained from the connected part $G^{\mathrm{con}}_r$ of the DMFT two-particle
Green's function. $G^{\mathrm{con}}_r$
can be computed by the DMFT impurity solver \cite{gunacker_worm_2015,gunacker_estim_2016}
and a subsequent combination of the spin components into channels $r=d,m$:
\begin{align}
G^{\mathrm{con}}_{d} & = G^{\mathrm{con}}_{\uparrow\uparrow\uparrow\uparrow}
                              + G^{\mathrm{con}}_{\uparrow\uparrow\downarrow\downarrow},\notag\\
G^{\mathrm{con}}_{m} & = G^{\mathrm{con}}_{\uparrow\uparrow\uparrow\uparrow}
                              - G^{\mathrm{con}}_{\uparrow\uparrow\downarrow\downarrow} \label{eq:su2_chan}.
\end{align}
The relation between $G^{\mathrm{con}}$ and $F$ is visualized in Fig.~\ref{fig:g_conn} and explicitly reads
\begin{equation}
F^{\omega\nu\nu'}_{r,lmm\pr l\pr} =  \sum_{nhn\pr h\pr}{[(\chi_0^{\omega})^{-1}]^{\nu\nu}_{lmn\pr h\pr} G^{{\mathrm{con~}}\omega\nu\nu'}_{r,h\pr n\pr nh} [(\chi^{\omega}_0)^{-1}]^{\nu'\nu'}_{hn m\pr l\pr}}. \label{eq:ffromgcon}
\end{equation}
The non-local, full vertex function $F^{\cvec{q}\nu'\nu}_r$ in Eq.~\eqref{eq:eta_imp} instead is obtained through the non-local version of the Bethe-Salpeter equation.

Usually, the Bethe-Salpeter starts from the irreducible vertex in a given channel $r$ and builds up ladder diagrams therefrom. This was also used in the first D$\Gamma$A calculations \cite{Toschi2007,Katanin2009}, whereas the dual fermion approach employed the local full vertex as a starting point \cite{Rubtsov2008,Hafermann2014a,Rubtsov12}. It turned out that the two approaches and the 1PI \cite{Rohringer2013}
actually construct the very same ladder diagrams, and only differ in how from this ladder, i.e., from the obtained $F^{\cvec{q}\nu'\nu}_r$, the self-energy is constructed, see e.g.\ Fig.~19 of \cite{RMPvertex}. Using the local full vertex as in DF instead of the irreducible vertex has the advantage that it does not suffer from vertex divergences \cite{Schaefer2013}, which is the reason why it is nowadays also employed in ladder D$\Gamma$A and why we employ it in AbinitioD$\Gamma$A.

 Following Ref.~\cite{Galler2017}, we hence rewrite Eq.~\eqref{eq:eta_imp} in the following compact form
\begin{equation}
\eta^{\cvec{q}\nu}_{r,lmm\pr l\pr} = \sum_{nh \nu\pr}(\vec{{\idmatrix}}_{lmhn} + \gamma^{\omega\nu\pr}_{r,lmhn}) \Big\{ \Big[ \Big( \idmatrix - \chi^{nl,\cvec{q}}_0 {F}^{\omega}_r - 2\beta^{-2}\chi^{\cvec{q}}_{0}{V}^{\kvec{q}}({\idmatrix} + \gamma^{\omega}_{r})\delta_{rd} \Big)^{-1}\Big]^{\nu\pr\nu}_{nhm\pr l\pr}-\idmatrix_{nhm\pr l\pr}\Big\},\label{eqn:eta_imp_mat}
\end{equation}
where $\idmatrix_{lmhn} = \delta_{ln}\delta_{mh}\delta_{\nu\nu\pr}$ and $\vec{\idmatrix}_{lmhn} = \sum_{\nu\pr} \idmatrix = \delta_{ln}\delta_{mh}$. Thus, $\eta^{\cvec{q}\nu}_{r}$ can be computed efficiently through a single matrix inversion and a consecutive multiplication with the three-leg quantity $(\vec{\idmatrix} + \gamma^{\omega\nu\pr}_{r})$ from the left. Note that the matrix that is inverted has a compound index consisting of one fermionic frequency and two orbitals for both, row and column [cf.~the indices $\{hn,\nu\pr \}$ and $\{m\pr l\pr,\nu \}$ after the inversion in Eq.~\eqref{eqn:eta_imp_mat}]. In the expression that is inverted in Eq.~\eqref{eqn:eta_imp_mat}, orbital and fermionic frequency indices have been omitted for clarity. For more details we refer the reader to Section \ref{Sec:compoundind} and Fig.~\ref{fig:g_matrix_1} (${F}^{\omega}$), Fig.~\ref{fig:g_matrix} ($\chi^{nl,\cvec{q}}_0$, $\chi^{\cvec{q}}_{0}$), and Figs.~\ref{fig:sum_left_nu},\ref{fig:u_gamma} ($\gamma^{\omega}$,${V}^{\kvec{q}}$).
Please also note that, by neglecting $\tilde{V}^{\kvec{k'}-\kvec{k}}$, the non-local Coulomb interaction $V^{\kvec{q}}$ needs to be added only in the density channel.

\subsection{Momentum-dependent susceptibilities}
\label{Sec:susc}
\noindent
With the AbinitioD$\Gamma$A program one can also compute momentum-dependent, physical DMFT susceptibilities. The susceptibilities $\chi_{r}^{\cvec{q}}$ in the density and magnetic channel $r\in\{d,m\}$ can be obtained from the three-leg vertices in Eqs.~\eqref{eq:gamma_loc_imp} and~\eqref{eq:eta_imp} according to

\begin{equation}
\begin{aligned}
\label{eq:susc}
\chi_{r,lmm'l'}^{\cvec{q}} &= \beta^{-2}\sum_{\nu\nu'}\chi_{r,lmm'l'}^{\cvec{q}\nu\nu'}\\ 
&=\chi_{0, lmm'l'}^{\cvec{q}} + \beta^{-2}\sum_{\substack{\nu\nu',nhh'n'}} \chi_{\substack{0, lmhn}}^{\cvec{q}\nu\nu} F^{\cvec{q}\nu\nu'}_{r, nhh'n'} \chi_{\substack{0, n'h'm'l'}}^{\cvec{q}\nu'\nu'} \\
&=\chi_{0, lmm'l'}^{\cvec{q}} + \beta^{-2} \sum_{\nu'h'n'} \eta_{r, lmh'n'}^{\cvec{q}\nu'} \chi_{\substack{0, n'h'm'l'}}^{\cvec{q}\nu'\nu'} + \beta^{-2} \sum_{\nu'h'n'} \gamma_{r, lmh'n'}^{\omega\nu'} \chi_{\substack{0, n'h'm'l'}}^{\cvec{q}\nu'\nu'}\\
&=\left(\chi_{0, lmm'l'}^{\omega}+\chi_{0, lmm'l'}^{\mathrm{nl},\cvec{q}}\right) + \beta^{-2} \sum_{\nu'h'n'} \eta_{r, lmh'n'}^{\cvec{q}\nu'} \chi_{\substack{0, n'h'm'l'}}^{\cvec{q}\nu'\nu'} + \beta^{-2} \sum_{\nu'h'n'} \gamma_{r, lmh'n'}^{\omega\nu'} \left(\chi_{\substack{0, n'h'm'l'}}^{\omega\nu'\nu'}+\chi_{\substack{0, n'h'm'l'}}^{\mathrm{nl},\cvec{q}\nu'\nu'}\right)\\
&=\chi_{r, lmm'l'}^{\omega} + \chi_{0, lmm'l'}^{\mathrm{nl},\cvec{q}} + \beta^{-2} \sum_{\nu'h'n'} \eta_{r, lmh'n'}^{\cvec{q}\nu'} \chi_{\substack{0, n'h'm'l'}}^{\cvec{q}\nu'\nu'} + \beta^{-2} \sum_{\nu'h'n'} \gamma_{r, lmh'n'}^{\omega\nu'} \chi_{\substack{0, n'h'm'l'}}^{\mathrm{nl},\cvec{q}\nu'\nu'},
\end{aligned}
\end{equation}
where
\begin{equation}
\label{eq:susc_loc}
\chi^{\omega}_{r,lmm\pr l\pr} = \beta^{-2}\sum_{\nu\nu\pr}\chi^{\omega\nu\nu\pr}_{r,lmm\pr l\pr}
\end{equation}
%
is the purely local DMFT susceptibility.
Please also note that the combined index $\cvec{q}$ in Eq.~\eqref{eq:susc} contains the momentum $\kvec{q}$ and the bosonic frequency $\omega$.
The usual magnetic and density susceptibilities are given by the spin-spin and charge-charge correlation functions
\begin{equation}
\chi_{m,ll\pr} = \left\langle \mathrm{T}_\tau \left(n_{l,\uparrow} - n_{l,\downarrow}\right)(\tau)\;\left(n_{l\pr,\uparrow} - n_{l\pr,\downarrow}\right)(0)\right\rangle
\label{eq:spincorr}
\end{equation}
\begin{equation}
\chi_{d,l l\pr} = \left\langle \mathrm{T}_\tau \left(n_{l,\uparrow} + n_{l,\downarrow}\right)(\tau)\;\left(n_{l\pr,\uparrow} + n_{l\pr,\downarrow}\right)(0)\right\rangle
\label{eq:chargecorr}
\end{equation}
\noindent
where $\mathrm{T}_\tau$ denotes the time-ordering operator.
These can be deduced from Eq.~\eqref{eq:susc} by choosing the orbital combinations $\chi^{\cvec{q}}_{r,lll\pr l\pr}$
and multiplying by a factor of 2. Finally, the physical susceptibilities, in units of $\mu_B^2/eV$,
can be readily obtained via an orbital summation:
\begin{equation}
\chi_{r} = 2\sum_{ll\pr} \chi^{\cvec{q}}_{r,lll\pr l\pr}
\label{eq:orbsum}
\end{equation}
\noindent
Note that for $r=m$ the above coincides with the magnetic susceptibility defined by

\begin{equation}
 \chi_m=g^2\sum_{ij}e^{i\kvec{q}(\kvec{R}_i-\kvec{R}_j)}\int d\tau e^{i\omega\tau}\left\langle\mathrm{T}_\tau S_{il}^z(\tau)S_{jl\pr}^z(0)\right\rangle \; ,
\end{equation}
\noindent
 when assuming a Land{\'e} factor $g=2$.
Let us emphasize again that the thus calculated susceptibility is the q-dependent DMFT susceptibility. For calculating distinct D$\Gamma$A susceptibilities a self-consistency or $\lambda$-correction \cite{Katanin2009} is needed.

\section{Installation and first AbinitioD$\Gamma$A run}
\label{Sec:instexample}
\subsection{Installation}
\label{Sec:installation}
\noindent
In the following we assume that all necessary dependencies are installed, namely:
\begin{itemize}
\item \verb=LAPACK= \cite{laug} library of version $3.8.0$ and above.
\item \verb=HDF5= \cite{hdf5} library of version $1.8.12$ and above.
\item \verb=h5py= \cite{collette_python_hdf5_2014} library of version $2.5.0$ and above.
\item \verb=numpy= \cite{numpy} library of version $1.9.1$ and above.
\end{itemize}
One way to conveniently obtain the code is via github (\verb=git= installation is required) or, alternatively, from the CPC repository:
\begin{verbatim}
$ git clone https://github.com/AbinitioDGA/ADGA.git
$ cd ADGA
\end{verbatim}
This will create the code directory \verb=ADGA= with all source files, documentation and test files included.
Before the program can be compiled, a configuration file called \verb=make_config= has to be created in the main \verb=ADGA= folder.
Here the necessary local path and environment variables are saved, on which the other Makefiles depend on.
Since this is strongly system-dependent, we give only a generic example which can also be found in\\
\verb=make_configs/make_config_cpc= :
\begin{verbatim}
F90       = mpifort
FPPFLAGS  = -DMPI
FFLAGS    = -O3
FINCLUDE  = -I/opt/hdf5-1.8.16_gcc/include/

LD        = $(F90)
LDFLAGS   = -I/opt/hdf5-1.8.16_gcc/include/ -L/opt/hdf5-1.8.16_gcc/lib/
LDFLAGS  += -lhdf5_fortran -lhdf5hl_fortran -llapack -lblas -limf
LDINCLUDE = -L/opt/hdf5-1.8.16_gcc/lib/
\end{verbatim}
The \verb=F*= variables describe the dependencies necessary for the compilation of the Fortran object files (\verb=*.o=)
while the \verb=LD*= variables describe the dependencies for the final linking process.
The compilation with MPI support\\
(\verb+FPPFLAGS = -DMPI+) and full optimization (\verb+FFLAGS = -O3+) is strongly recommended. Compilation without
MPI support however is still available by simply using a non-MPI compiler (e.g., \verb=gfortran=) and leaving the \verb=FPPFLAGS= variable empty. The rest of the variables contain
absolute paths to the respective, mandatory libraries in combination with their standard linking variables (\verb=-lhdf5_fortran= etc.).

After successful creation of this file the AbinitioD$\Gamma$A program can be compiled by simply executing
\begin{verbatim}
$ make
\end{verbatim}
The compilation process automatically creates the subdirectory \verb=bin= containing the executables \verb=abinitiodga=
and \verb=setupvertex= (see Fig.~\ref{fig:abinitdga_flow_new}).

\subsection{Setting up a calculation}
\label{Sec:Setup}
\noindent
In this section we will illustrate the steps needed to perform an AbinitioD$\Gamma$A calculation. The
necessary input data, obtained from \verb=w2dynamics= \cite{w2dynamicsCPC} and a Wannier90 Hamiltonian,
is for this test case already included in the github repository (in the subfolder \verb+srvo3-testdata+):
\begin{itemize}
\item \verb=srvo3-1pg.hdf5= (\verb=w2dynamics= DMFT output file)
\item \verb=srvo3-2pg.hdf5= (\verb=w2dynamics= worm-sampling vertex output file)
\item \verb=srvo3_k20.hk=   (\verb=wien2wannier= Wannier Hamiltonian with \verb=20x20x20= k-points)
\end{itemize}
\noindent
The provided files contain only data within a reduced range of Matsubara frequencies (frequency box)
to speed up the test calculation.

As mentioned in Sec.~\ref{Sec:equations} we first have to symmetrize the spin-components of our vertex.
Additionally, for the case of locally degenerate (equivalent) orbitals, also orbital-components can be symmetrized (option (o) below).
This is done by executing the program \verb=setupvertex= with the following user options:
\begin{verbatim}
$ cd srvo3-testdata
$ ../bin/setupvertex
Number of inequivalent atoms: 1
Vertex file: srvo3-2pg.hdf5
Number of correlated bands for inequivalent atom 1: 3
Outputfile for symmetrized Vertex: srvo3-2pg-symmetrized.hdf5

SU2 symmetry only (s) or SU2 AND orbital symmetry (o)?: o
\end{verbatim}

\noindent
This way, the symmetrized vertex has been saved under the specified file name \verb=srvo3-2pg-symmetrized.hdf5=. Now all necessary input files are prepared and we are ready to use the main AbinitioD$\Gamma$A program \verb=abinitiodga=. For the latter we still need to setup a config file. For the sake of brevity in this section
we use the provided config file listed in \ref{app:configself} and save it as \verb=case.conf= (The configuration file
will be explained in more detail in Sec.~\ref{sec:adv-user-opt}).
The code can now be run by executing
\begin{verbatim}
$ mpirun -np $NCORES ../bin/abinitiodga case.conf
\end{verbatim}
where \verb=$NCORES= refers to the number of cores used.
While running, the program writes runtime information into a log file called \verb=out=.
The output data instead is written into a custom HDF5 file. Depending on the run options the structure of the latter changes slightly.
All possible output datasets are listed in detail in the provided \verb=README.pdf=. The data of the HDF5 output file can be extracted via
the h5py library in Python. We provide several scripts with a detailed documentation of all steps in \verb=documenation/scripts=.
The reference results (including prepared plot scripts) can be found in \ref{app:results}.

\subsection{SrVO$_3$ as test example}
\label{Sec:example}
\noindent
In this section we discuss the results of the previously set-up calculation for SrVO$_3$.
The reduced range of Matsubara frequencies used in the previously set-up test calculation is too small
to yield fully converged results. In the following sections we show the results calculated
for much larger frequency box sized, and discuss their physical interpretation.

Our target material is a strongly correlated, metallic transition metal oxide with a cubic perovskite crystal structure whose low-energy physics is dominated by its degenerate vanadium $t_{2g}$ states. Experimentally, several manifestations of electronic correlations have been observed in SrVO$_3$: photoemission spectroscopy \cite{Sekiyama2004} and specific heat measurements \cite{PhysRevB.58.4372} find a mass enhancement of a factor of two compared to band-theory, the spectral function exhibits a satellite feature, i.e.~a Hubbard band, below the quasiparticle peak \cite{Sekiyama2004,PhysRevB.52.13711,PhysRevB.80.235104}, and at closer look a kink in the energy-momentum dispersion becomes visible \cite{Nekrasov05a,Aizaki12,Held13}. Theoretically these phenomena have extensively been studied within various methods for strongly correlated electron systems. Indeed SrVO$_3$ has become a textbook example and testbed material in this field. Nonetheless, several physical aspects of this material are still under discussion and have recently been re-investigated with new, post-DMFT techniques. Calculations that include the dynamical nature of the screened Coulomb interaction suggest that a sizable part of the mass enhancement in SrVO$_3$ may originate from plasmon excitations \cite{Casula2012,Tomczak2012,PhysRevB.88.235110,Tomczak2014,Boehnke2016}. Furthermore, screened exchange contributions to the self-energy---that are caused by non-local interactions $V^\kvec{q}$---have been shown to compete with the mass enhancement from dynamical correlations \cite{Tomczak2014,Boehnke2016,Miyake13}.
Besides this academic interest, SrVO$_3$ has potential for technological applications, e.g., as an electrode material \cite{ADMA:ADMA201300900}, Mott transistor \cite{PhysRevLett.114.246401} or transparent conductor \cite{Zhang2016}.
Hence, SrVO$_3$ is a suitable target material for illustrating the capabilities and usage of our new AbinitioD$\Gamma$A algorithm.

\begin{figure*}[t!]
\begin{minipage}{16.5 cm}
\includegraphics[clip=true,width=16.5cm]{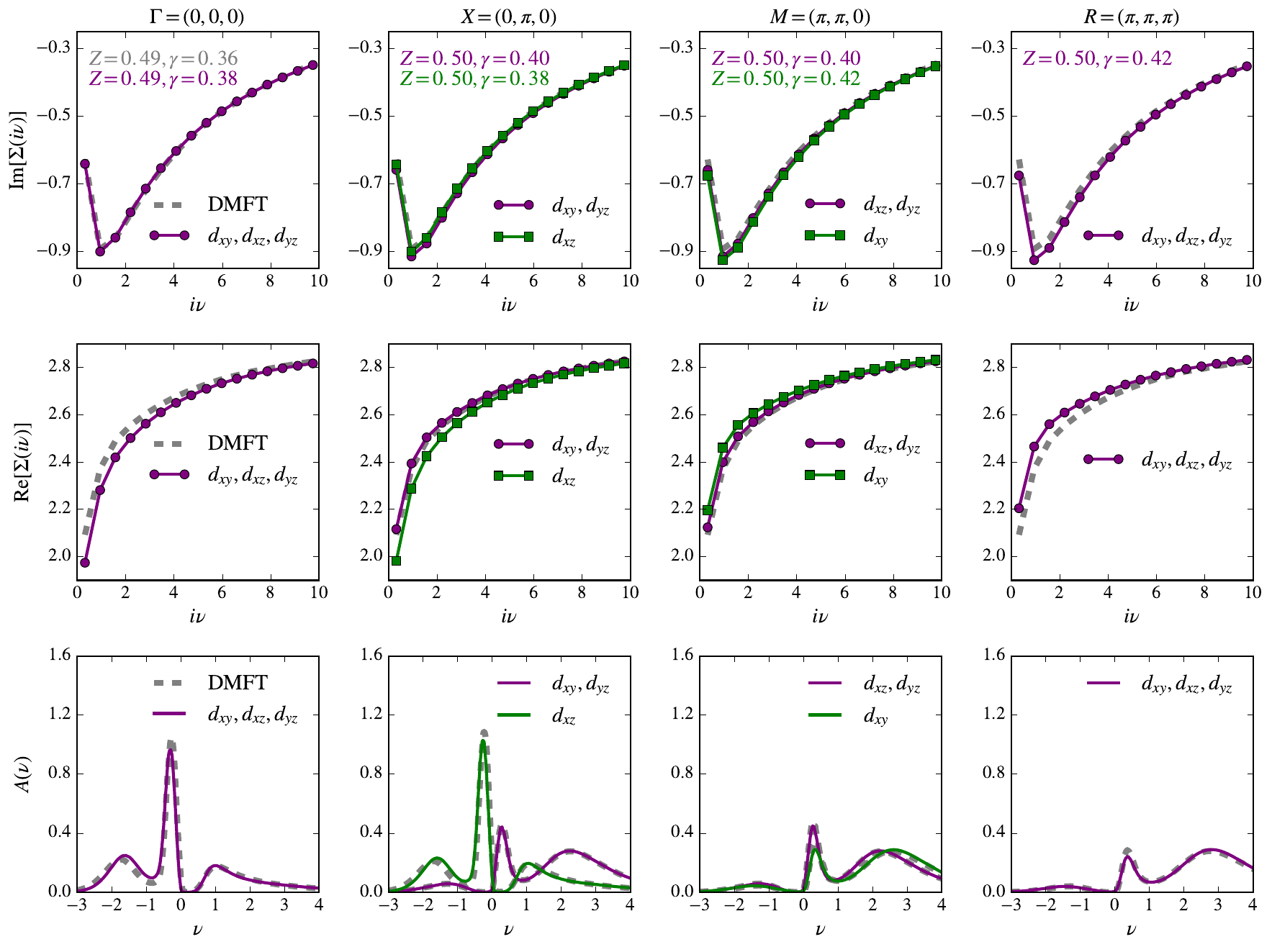}
\caption{AbinitioD$\Gamma$A self-energies (top row: imaginary part, middle row: real part) and spectral functions (bottom row) for selected k-points in the Brillouin zone ($\Gamma$-point: first column, $X$-point: second column, $M$-point: third column, $R$-point: fourth column).
In the two top panels, symbols indicate results for a small frequency box ($\#\omega/2=60$), while lines depict results for a large box ($\#\omega/2=200$): Congruence shows that the calculation is converged
with respect to the number of frequencies.}
\label{fig:siw_multiplot}
\end{minipage}
\end{figure*}

\subsubsection{DFT+DMFT}
\noindent
For the present example case, the Wien2k program package~\cite{blaha_wien2k,Schwarz2003} was used to
perform the DFT computation, and the wien2wannier interface~\cite{wien2wannier}
and wannier90~\cite{wannier90} to construct the Wannier Hamiltonian $H_W^{\mathbf{k}}$.
The subsequent DMFT calculation was done with w2dynamics~\cite{w2dynamicsCPC,w2dynamics,Wallerberger16}
and its continuous-time quantum Monte Carlo (CTQMC)
algorithm in the hybridization expansion (CT-HYB)~\cite{Werner2006,Werner2006a,Gull2011a}.
The AbinitioD$\Gamma$A program is adapted to the output formats of the described program packages.
Let us mention here, however, that it is possible to reformat any DMFT code output to conform to the AbinitioD$\Gamma$A input standards.

\subsubsection{AbinitioD$\Gamma$A self-energies}
\noindent
The results shown here, have been obtained \cite{JPSJ-XXX} using only local Coulomb
interactions in the Kanamori parametrization with $U=5$eV, $V=3.5$eV and $J=0.75$eV
at an inverse temperature $\beta=10$eV$^{-1}$.
The calculation was done on a $k$- and $q$-grid of $20\times 20\times 20$ points.

The AbinitioD$\Gamma$A self-energies and corresponding spectral functions for four high-symmetry k-points are shown in Fig.~\ref{fig:siw_multiplot}. In the two top panels, the real and imaginary part of the self-energy are displayed in color (green and violet) while the momentum-independent DMFT self-energy is shown in gray. Beside its k-dependence, the AbinitioD$\Gamma$A self-energy $\Sigma^{\cvec{k}}_{mm\pr}$ is also orbital-dependent: Different colors (green and violet) in Fig.~\ref{fig:siw_multiplot} refer to diagonal components $\Sigma_{mm}$ of different orbitals $m=d_{xy},d_{xz},d_{yz}$. In general, also orbital-offdiagonal ($m\neq m\pr$) components of the self-energy can arise if allowed by symmetry. In this SrVO$_3$ example case however they are very small, so that we show only orbital-diagonal components.

From a physical point of view, the top panel of Fig.~\ref{fig:siw_multiplot} shows that at low energies the imaginary part of the self-energy on the Matsubara axis is---for all orbitals and k-points---very close to the DMFT result.
As a result the scattering rate $\gamma=-\Im\Sigma(i\nu\rightarrow 0)$ and the quasi-particle weight $Z_\mathbf{k}$ do not depend significantly on the momentum, which is a quite common finding within post-DMFT methods \cite{Galler2017,Tomczak2014,jmt_pnict,jmt_sces14,Schaefer2015}---at least in three spatial dimensions.
The real-part of the self-energy instead shows larger deviations from the DMFT result. Indeed, at low energies the difference between AbinitioD$\Gamma$A and DMFT reaches 200meV.

The bottom panels of Fig.~\ref{fig:siw_multiplot} show the AbinitioD$\Gamma$A spectral
functions. They were obtained by analytically continuing the Matsubara Green's function
to the real-frequency axis, using the maximum entropy method \cite{Jarrell1996,Sandvik98b,w2dynamicsCPC}. As expected from the analysis
of the self-energies, in the spectral functions we see signatures of reduced correlation
effects compared to the DMFT results (in gray): the quasi-particle peaks move very slightly
away from the Fermi level while Hubbard bands are displaced towards it.
For a more detailed discussion and physical interpretation of the results please refer to Ref.~\cite{JPSJ-XXX}.

\subsubsection{DMFT susceptibilities}
\begin{figure}
\centering
\includegraphics[width=0.48\textwidth]{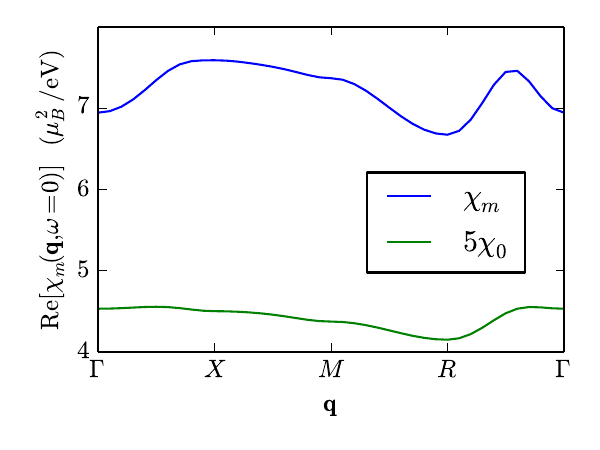}
\includegraphics[width=0.48\textwidth]{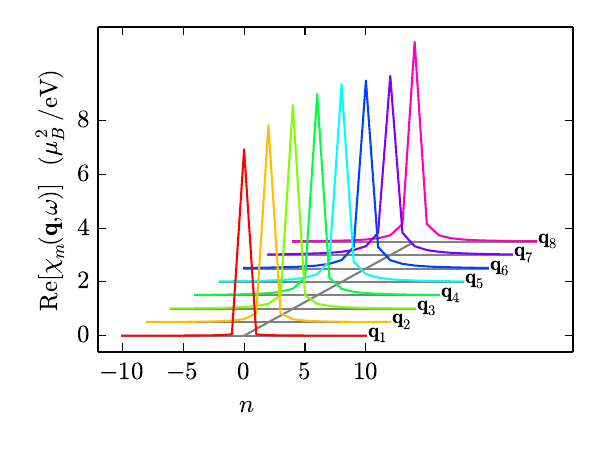}
\caption{\label{fig:chi_qw} Magnetic susceptibilities of SrVO$_3$. Left: real part of the magnetic susceptibility $\chi_m$ at zero frequency, along a q-path connecting high-symmetry points in momentum space.
For comparison, we also show the particle-hole bubble, $\chi_0$, scaled by a factor of 5.
Right: real part of the dynamical magnetic susceptibility $\chi_m(i\omega_n)$ of Eq.~\ref{eq:orbsum} as a function of bosonic Matsubara frequencies $\omega_n=2\pi n/\beta$ for 8 selected q-points:
${\mathbf q}_1$ = (0,0,0),
${\mathbf q}_2$= ($\pi/2$,0,0),
${\mathbf q}_3$ = ($\pi$,0,0),
${\mathbf q}_4$ = ($\pi$,$\pi/2$,0),
${\mathbf q}_5$ = ($\pi$,$\pi$,0),
${\mathbf q}_6$ = ($\pi$,$\pi$,$\pi/2$),
${\mathbf q}_7$ = ($\pi$,$\pi$,$\pi$),
${\mathbf q}_8$ = ($\pi/2$,$\pi/2$,$\pi/2$). Please note that curves have been offset for better visibility.}
\end{figure}

\noindent
Results for the {\it static} DMFT susceptibility of SrVO$_3$ are shown in the left panel of Fig.~\ref{fig:chi_qw}, summed over all orbital contributions ($\chi_r = 2 \sum_{lm} \chi_{r, llmm}$). There is only a weak momentum-dependence of the magnetic susceptibility in the high-temperature paramagnetic phase of SrVO$_3$, but vertex corrections (susceptibilities beyond $\chi_0$ of Eq.~\eqref{eq:q_bubble}) strongly enhance the susceptibility by a factor of seven.

Indeed only if vertex corrections are
taken into account the susceptibility agrees with the experimental value \cite{Lan2003,Burzo1996}. For example, Ref.~\cite{Lan2003} reports $9.9\mu_B^2/{\rm eV}^{-1}$ at 100K and $8.7\mu_B^2/{\rm eV}^{-1}$ at 300K which, taken the temperature dependence into account, well agrees with our value of $\approx\!\!7\mu_B^2/{\rm eV}^{-1}$ at 1160K (for the conversion of units note that $1\mu_B^2/{\rm eV} = 3.2327\times 10^{-5} {\rm emu}/ {\rm mol}$).

\noindent
We are also able to produce the dynamical susceptibilities
shown in the right panel of Fig.~\ref{fig:chi_qw}. These still need to be analytically continued to the physical (real frequency) dynamical susceptibilities.
The susceptibilities on the Matsubara axis shown in Fig.~\ref{fig:chi_qw} allow however for a better intercomparison and test case without the perils of analytical continuation, e.g., by the maximum entropy method.

\section{Detailed user options}
\label{sec:adv-user-opt}
\begin{figure}[h]
\centering
\includegraphics[width=14cm]{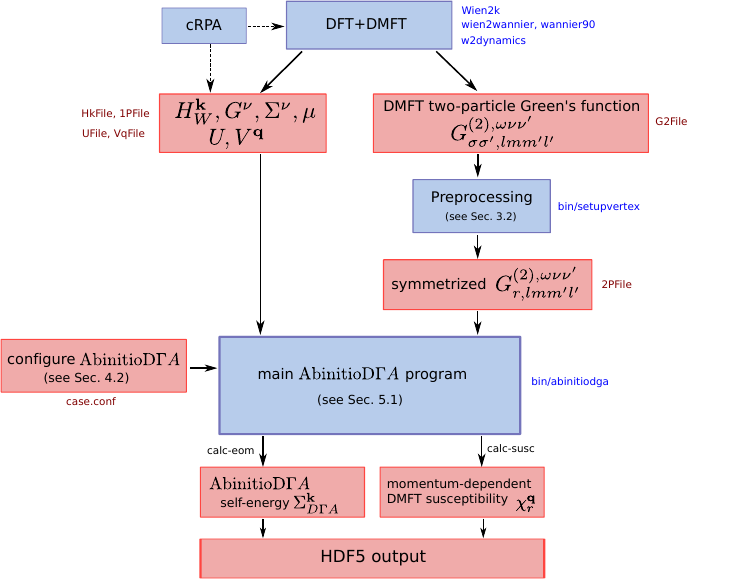}
\caption{Flow diagram of the AbinitioD$\Gamma$A algorithm. Programs are indicated as blue boxes with the program names on the right. The input/output data files instead are indicated in red.}
\label{fig:abinitdga_flow_new}
\end{figure}

\noindent
In this section we go into more detail regarding the program flow, run options and the effect of different parameters on the calculation (all from a user's point of view).
The general program flow with its interfaces to other program packages is illustrated in Fig.~\ref{fig:abinitdga_flow_new}.

\subsection{DFT+DMFT input}
\label{dftdmft}
\noindent
The usual starting point of AbinitioD$\Gamma$A is a converged DFT+DMFT calculation for the material under investigation.
It provides the Wannier Hamiltonian $H_W^{\kvec{k}}$ in the format of the current version of the \verb=convham= program of Wien2k (version $18.2$)
and the local DMFT one-particle Green's function $G^{\nu}$,
self-energy $\Sigma^{\nu}$ and double counting $\Sigma_{DC}$ as well as the chemical potential $\mu$
in an HDF5 file of the structure of the w2dynamics output format in its current version $1.0.0$. For other DFT and DMFT programs this HDF5 file needs to be generated before using AbinitioD$\Gamma$A

The local and non-local Coulomb interaction $U$ and $V^{\mathbf{q}}$ can also be obtained
\textit{ab initio} by using the constrained random phase approximation (cRPA)~\cite{cRPA_2004,cRPA_wannier_2008}.

The four-index $U$ can either be stored in plain text file or be constructed directly
by specifying the interaction parameters and interaction type in the config file of the main AbinitioD$\Gamma$A program.
The purely nonlocal interaction $V^{\mathbf{q}}$, on the other hand, is stored in a separate HDF5 file whose format is described in the file \verb=README.pdf=.

After convergence of the DFT+DMFT cycle, the full DMFT two-particle Green's function
$G^{(2)}$, or its connected part $G^{\mathrm{con}}$, are computed.
Since in a multi-orbital impurity the local interaction is usually not of density-density type,
this requires a worm sampling technique \cite{gunacker_worm_2015,gunacker_estim_2016}.

Subsequently, the spin components of the two-particle Green's function are combined into the density ($d$) and magnetic ($m$) channel by using the program \verb=setupvertex=, cf.~Fig.\ref{fig:abinitdga_flow_new}.

\subsection{Configuration file} \label{sec:main_prog}
\label{Sec:mainAbinitioprogram}
\noindent
The AbinitioD$\Gamma$A program \verb=abinitiodga= uses a free-format configuration file.
A full documentation of the latter can be found in \verb=documentation/configspec= and \verb=documentation/README/README.pdf= in the code repository.
Here we explain the most important parameters based on
the config file corresponding to the example case in Section \ref{Sec:example}.

The config file is structured into several sections marked by square brackets.
In the first section, the user needs to specify some general parameters:
\begin{verbatim}
[General]
calc-susc = T # calculate the momentum-dependent susceptibilities: (T)rue / (F)alse
calc-eom  = T # calculate the dga-selfenergy via the equation of motion: (T)rue / (F)alse

# number of positive f/b frequencies of the vertex
N4iwf = -1 # full box
N4iwb = -1 # full box

# Number of atoms
NAt = 1

# Wannier Hamiltonian
HkFile = srvo3_k20.hk

k-grid = 20 20 20 # Wannier Hamiltonian and eom momentum grid
q-grid = 20 20 20 # Grid for susc, and q-sum in eom
\end{verbatim}
Here, we first specified that we want to calculate the DMFT susceptibility (\verb=calc-susc=) and D$\Gamma$A self-energies (\verb=calc-eom=).
The fermionic and bosonic frequency box sizes of the vertex are given by the parameters \verb=N4iwf= and \verb=N4iwb=. These two parameters can be used to check the convergence of the calculation with respect to the frequency box size: \verb=N4iwf= $=-1$ employs the maximum frequency box (defined by the previous CT-QMC calculation), while one can also choose a smaller number of frequencies by explicitly setting, e.g., \verb=N4iwf= $=20$.
The parameter \verb=NAt= specifies the number of correlated atoms in the calculation and
the Wannier Hamiltonian $H_{W}^\svek{k}$ in the reducible Brillouin zone (in the format of wien2wannier) is read from the file \verb=HkFile=.
\verb=k-grid= specifies the number of k-points in each direction (the \verb=k-grid= must coincide with the one of the Wannier Hamiltonian). \verb=q-grid= instead controls the momentum grid convergence and only affects
the self-energies due to the internal momentum sum (Please note that only an integer divisor for each direction is allowed, e.g., \verb=4 5 10= in the case of a k-grid of \verb=20 20 20=).

Next, one needs to define, for each correlated atom \verb=NAt=, the number of orbitals and the interaction type:
\begin{verbatim}
[Atoms]
[[1]]
Interaction = Kanamori # interaction type
Nd = 3 # number of d-orbitals
Np = 0 # number of p-orbitals
Udd = 5.0 # intra-orbital interaction
Vdd = 3.5 # inter-orbital interaction
Jdd = 0.75 # Hund's coupling
\end{verbatim}
For this configuration, the program automatically generates a Kanamori interaction matrix with the given
interaction parameters \verb=Udd=, \verb=Vdd= and \verb=Jdd=.
Alternatively, it is possible to provide a full four-index $U_{lm\pr ml\pr}$ in form of a plain text file \verb=UFile= (examples are provided in \verb=documentation/examples=).
Besides the local interaction used in DMFT, a completely non-local interaction ($V^\svek{q}$, with $\sum_\svek{q} V^\svek{q} = 0$)
can be specified in \verb=VqFile=. The latter is a HDF5 file that contains only the non-zero spin-orbital components of $V^\kvec{q}$ in the form of a lookup table.

Finally, we provide the one-particle data in the usual \verb=w2dynamics= HDF5 output format.
The two-particle data, on the other hand, uses a special HDF5 file format (described in Fig.~\ref{fig:g4iw_sym} below) which is
automatically generated by the program \verb=setupvertex= described in Sec.~\ref{Sec:Setup}.
\begin{verbatim}
[One-Particle]
# w2dynamics DMFT output file
1PFile = srvo3-1pg.hdf5

[Two-Particle]
# symmetrized vertex
2PFile = srvo3-2pg-symmetrized.hdf5
# legacy option
# 0: full two-particle Greens function, including disconnected parts
vertex-type = 0
\end{verbatim}

\subsection{Output}
\noindent
AbinitioD$\Gamma$A uses HDF5 for its main output.
The name of the output file is generated automatically and contains the time stamp of the start of the calculation
in order to make its name unique and identifiable (example: \verb=adga-20180904-022233.615-output.hdf5=).

\begin{figure}[h]
\centering
\includegraphics[width=9cm]{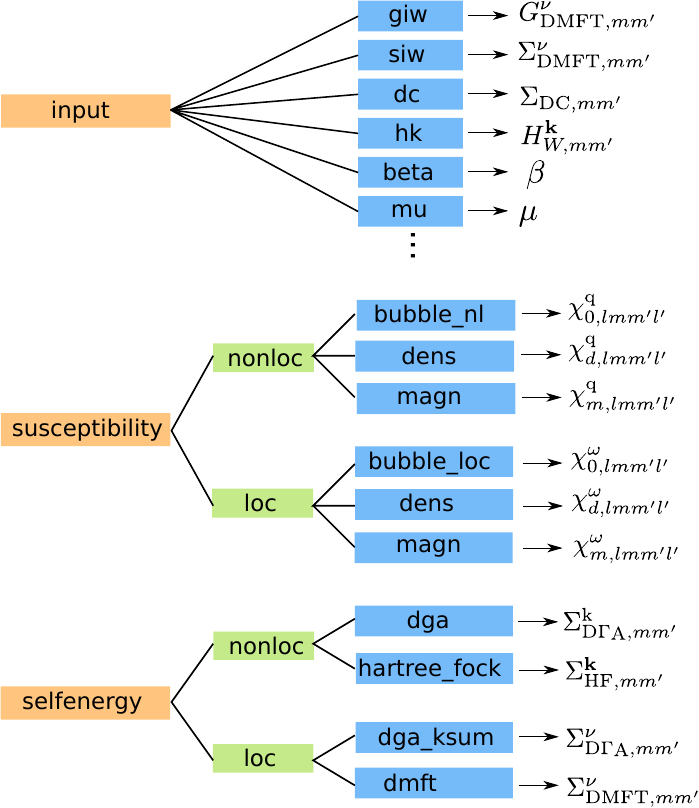}
\caption{Group structure of the AbinitioD$\Gamma$A HDF5 output.}
\label{fig:output}
\end{figure}

At the top level, the file contains three groups (see Fig.~\ref{fig:output}):
\begin{enumerate}
  \item \verb=input= consists of several datasets, in which the DFT+DMFT input data (but not the two-particle Green's function) are stored.
    This is done merely for convenience, so as to simplify, e.g., comparisons of the DMFT self-energy (stored in \verb=input/siw=) with the AbinitioD$\Gamma$A self-energy.
  \item \verb=susceptibility= contains groups for both the local (\verb=loc=) and non-local (\verb=nonloc=) DMFT-susceptibility of Eqs.~\eqref{eq:susc} and~\eqref{eq:susc_loc}
    in the density and magnetic channel.
  \item \verb=selfenergy= contains the AbinitioD$\Gamma$A self-energy $\Sigma^{\cvec{k}}_{\mathrm{D}\Gamma\mathrm{A}}$
    of Eq.~\eqref{eq:eom_final_imp} in the subgroup \verb=nonloc/dga=.
    However, the latter does not contain the non-local Hartree-Fock term $\Sigma_{\mathrm{HF~}}^{\kvec{k}}$
    of Eq.~\eqref{eq:eom_hartree} which is stored separately in \verb=nonloc/hartree_fock=.
    The subgroup \verb=loc/dga_ksum= contains the local ($\kvec{k}$-summed) AbinitioD$\Gamma$A self-energy
    $\Sigma^\nu_{\mathrm{D}\Gamma\mathrm{A}}=\sum_{\kvec{k}}\Sigma^{\cvec{k}}_{\mathrm{D}\Gamma\mathrm{A}}$.
    Furthermore, \verb=loc/dmft= contains the DMFT self-energy obtained through the local version
    of the equation of motion,
    i.e.~$\Sigma^\nu_{{\mathrm{DMFT}}}=-\beta^{-1}\sum_{\omega}U\gamma^{\omega\nu}_d G^{\cvec{\nu-\omega}}+\Sigma_{\mathrm{HF}}^U$ (with the local Hartree-Fock term $\Sigma_{\mathrm{HF}}^U$ included).
    Up to statistical fluctuations, the latter coincides with the 'original' DMFT self-energy stored in \verb=input/siw=.
    A cross-check of \verb=input/siw= with \verb=selfenergy/loc/dmft= is always recommended.
\end{enumerate}
The file \verb=README.pdf= in the code repository contains a complete listing of all groups and datasets of the output file.
Most conveniently they can be accessed and plotted in python by \verb=h5py= and the \verb=matplotlib=. Exemplary python scripts are provided in \verb=documentation/scripts=.
\\[2\baselineskip]\noindent

\newpage
\clearpage
\section{Program structure and algorithmic details}
\label{sec:programalgo}

\noindent
In this section we are going to switch from the user's to the developer's point of view. We are first going to introduce
the internal program flow in Sec.~\ref{sec:programstructure} and then go into detail about the internal storage layout
and the resulting characteristics of the matrix operations in Sec.~\ref{Sec:algdetails}.

\subsection{Program structure}
\label{sec:programstructure}
\noindent
The program is set up in three distinct steps shown in Table \ref{tab:programflow}. First we initialize all the necessary interfaces
and read in (or calculate) all the necessary variables which are needed throughout the program. After initializing the MPI and HDF5 interface
we read and check the config file (described in detail in Sec.~\ref{sec:adv-user-opt}).
Here we check for inconsistencies in the provided data
and the provided run-options. If no errors are detected we read in the data necessary for the construction of the Green's function $G^{\cvec{k}}_{hh\pr}$ \eqref{eq:g_nonloc}
and the interaction matrix $U_{mlnh}$. At the end of this first step the work load is distributed to all the MPI processes (if compiled with MPI) and we can initialize the HDF5 output file.

\begin{table}[h]
\centering
\begin{tabular}{|ll|lll|}
\hline
\multicolumn{3}{|l}{\textbf{}}  & \multicolumn{1}{l}{\textbf{function name}}     & \multicolumn{1}{l|}{\textbf{file}}            \\
\hline
\hline
\multicolumn{3}{|l}{initialize MPI}  & mpi\_initialize     & mpi\_org.F90            \\
\multicolumn{3}{|l}{initialize HDF5 interface}  & init\_h5            & hdf5\_module.f90        \\
\multicolumn{3}{|l}{read \& check config file}    & read\_config     & config\_module.f90      \\
\multicolumn{3}{|l}{}    & check\_config       & config\_module.f90                         \\
\multicolumn{3}{|l}{read one-particle info}  & read\_siw           & hdf5\_module.f90        \\
\multicolumn{3}{|l}{}  & read\_giw or create\_giw           &     hdf5\_module.f90                    \\
\multicolumn{3}{|l}{}  & ...                 &                         \\
\multicolumn{3}{|l}{create interaction matrix $U_{mlnh}$}  & read\_u or create\_u & interaction\_module.f90 \\
\multicolumn{3}{|l}{calculate\;\;$n^\mathbf{k}$}  & get\_nfock & one\_particle\_quant\_module.f90 \\
\multicolumn{3}{|l}{distribute MPI work load}  & mpi\_distribute     & mpi\_org.F90            \\
\multicolumn{3}{|l}{initialize HDF5 output file}  & init\_h5\_output    & hdf5\_module.f90        \\
\hline
\hline
\multicolumn{3}{|l}{\textbf{LOOP: bosonic frequency -- $i\omega_n$}}  &                     &                 \\\cline{1-5}
{} & \multicolumn{2}{|l}{calculate\;\;$\chi_0^\omega$ \eqref{eq:bubble_loc}}  &   get\_chi0\_loc        &              one\_particle\_quant\_module.f90  \\
{} & \multicolumn{2}{|l}{read $i\omega_n$ slice of $G_r^{\mathrm{con}\;\omega\nu\nu\prime}$ \eqref{eq:su2_chan}}  & read\_vertex        & hdf5\_module.f90        \\
{} & \multicolumn{2}{|l}{calculate\;\;$\chi_0^\omega F_r^\omega$ \eqref{eq:ffromgcon}}  &                     &                 \\
{} & \multicolumn{2}{|l}{calculate\;\;$\gamma_r^\omega$ \eqref{eq:gamma_loc_imp}}  &                     &                 \\
{} & \multicolumn{2}{|l}{calculate\;\;$\gamma_r^\omega \chi_0^\omega$ \eqref{eq:susc}} & calc\_chi\_qw & susc\_module.f90                             \\
{} & \multicolumn{2}{|l}{calculate\;\;$\Sigma^\mathrm{DMFT}$ (for comparison)}  & calc\_eom\_dmft & eom\_module.f90 \\\cline{2-5}
{} & \multicolumn{2}{|l}{\textbf{LOOP: transferred momentum -- $\mathbf{q}$}} & &  \\\cline{2-5}
{} & {} & optional: read $V(\mathbf{q})$  & read\_vq & interaction\_module.f90 \\
{} & {} & calculate\;\;$\Sigma^\mathrm{HF}$ \eqref{eq:eom_hartree} & calc\_eom\_static & eom\_module.f90 \\
{} & {} & calculate\;\;$\chi_0^\cvec{q}$ \eqref{eq:q_bubble} & accumulate\_chi0 & one\_particle\_quant\_module.f90 \\
{} & {} & calculate\;\;$\chi_0^\cvec{nl} = \chi_0^\cvec{q} - \chi_0^\omega$ \eqref{eq:bubble_nl} &  &  \\
{} & {} & calculate\;\;$\gamma_r^\omega \chi_0^\cvec{nl}$ \eqref{eq:susc} & calc\_chi\_qw & susc\_module.f90 \\
{} & {} & calculate\;\;$\gamma^\cvec{q}_r \eqref{eq:gamma_q_imp}$ & {} &  \\
{} & {} & calculate\;\;$\eta^\cvec{q}_r$ \eqref{eq:eta_imp}& {} &   \\
{} & {} & calculate\;\;$\eta^\cvec{q}_r\chi_0^\cvec{q}$ \eqref{eq:susc} & calc\_chi\_qw &susc\_module.f90  \\
{} & {} & calculate\;\;$\Sigma^\mathrm{D\Gamma A} \eqref{eq:eom_final_imp}$ & calc\_eom\_dynamic & eom\_module.f90 \\
\hline
\hline
\multicolumn{3}{|l}{gather (and/or sum) data from MPI processes}  & {}           &                         \\
\multicolumn{3}{|l}{output data to HDF5 output file}  & {output\_eom}           &    hdf5\_module.f90                     \\
\multicolumn{3}{|l}{}  & {output\_chi\_loc}           &             hdf5\_module.f90            \\
\multicolumn{3}{|l}{}  & {output\_chi\_qw}           &              hdf5\_module.f90           \\
\multicolumn{3}{|l}{}  & {...}           &                         \\
\hline
\end{tabular}
\label{tab:programflow}
\caption{Internal program flow of the AbinitioD$\Gamma$A program. The first column lists, step-by-step, the instructions while the second and third column
shows the corresponding functions and where to find them. An empty function name implicates that the task is performed in the main file (main.F90).}
\end{table}
\noindent
In the second step we perform the two computationally heavy main
loops (bosonic frequency $i\omega_n$ and transferred momentum $\mathbf{q}$) where we calculate, step-by-step, the objects which are required
for the momentum-dependent susceptibilities $\chi_{r,lmm\pr l\pr}^\cvec{q}$ \eqref{eq:susc} and the D$\Gamma$A self-energy
$\Sigma^{\cvec{k}}_{\substack{{\mathrm{D}}\Gamma{\mathrm{A}}\\ mm\pr}}$ \eqref{eq:eom_final_imp}. Thirdly,
we gather all the data, perform all necessary sums and write it to the HDF5 output file.

\subsection{Algorithmic details}
\label{detail}
\label{Sec:algdetails}

\subsubsection{Storage of the DMFT two-particle Green's function} \label{sec:store_g4iw}
\label{Sec:storageDMFT2PGF}
\noindent
The DMFT two-particle Green's function $G^{(2)}$, which can be measured in CT-HYB,
is a very large quantity and needs a lot of storage capacity.
In its most general form $G^{(2)}$ has four orbital indices $lmm\pr l\pr$ and four spin indices $\sigma_1\sigma_2\sigma_3\sigma_4$,
and it depends on three Matsubara frequencies $\omega\nu\nu'$: $G^{(2)\, \omega\nu\nu'}_{\sigma_1\sigma_2\sigma_3\sigma_4, lmm\pr l\pr}$.
However, the orbital and spin degrees of freedom are restricted by the symmetries of the local DMFT impurity problem.
The Kanamori parameterization of interactions allows only for orbital combinations with pairwise identical orbitals:
($iijj$, $ijij$, $ijji$). Furthermore, in the SU(2)-symmetric and paramagnetic case,
also the spin degrees of freedom are reduced.
Hence, many spin-orbital components of $G^{(2)\, \omega\nu\nu'}_{\sigma_1\sigma_2\sigma_3\sigma_4, lmm\pr l\pr}$ are actually zero.
Thus, the amount of storage for $G^{(2)}$ can be massively reduced by storing only its non-zero spin-orbital components in the file \verb=G2File= (see Fig.~\ref{fig:abinitdga_flow_new}).
This reduces the required storage space by a factor of $\frac{[2(\#o)]^4}{6[3(\#o)^2-2(\#o)]}\approx 10$
for the example calculation in Section \ref{Sec:example}.

\begin{figure}[h]
\centering
\includegraphics[width=5cm]{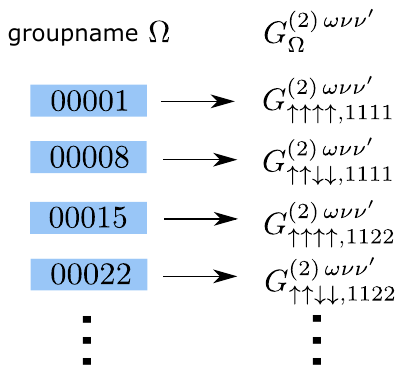}
\caption{Structure of the {\tt{G2File}}, which contains the two-particle impurity Green's function $G^{(2)}$.
Each group---named by the combined index $\Omega$---contains the corresponding
non-zero spin-orbital component of $G^{(2)}$.
Shown are the first four groups in the SrVO$_3$ example file.
Note that many spin-orbital combinations do not exist and are hence not stored,
e.g., $\Omega=2$ which corresponds to $G^{(2)\, \omega\nu\nu'}_{\uparrow\uparrow\uparrow\downarrow,1111}$.}
\label{fig:g4iw_file}
\end{figure}

\paragraph*{Group structure of the \tt{G2File}} The non-zero spin-orbital components of $G^{(2)}$
are stored in the HDF5 file \verb=G2File= in the form of a "lookup-table".
This means that the band and spin indices of $G^{(2)\, \omega\nu\nu'}_{\sigma_1\sigma_2\sigma_3\sigma_4, lmm\pr l\pr}$
are translated into a single index $\Omega$ through a unique transformation:
\begin{equation}
\sigma_1\sigma_2\sigma_3\sigma_4, lmm\pr l\pr \leftrightarrow \Omega . \label{ind_trafo}
\end{equation}
The index $\Omega$ is then used to store the non-zero spin-orbital components of $G^{(2)}$ in \verb=G2file=.
That is, the index $\Omega$ is the name of the groups in \verb=G2File= containing the corresponding
non-zero spin-orbital component of $G^{(2)}$. Thus, \verb=G2File= contains as many groups
as there are non-zero spin-orbital components in $G^{(2)}$.
For example, for SrVO$_3$ in a paramagnetic $t_{2g}$ setup the structure of the \verb=G2File=
is shown in Fig.~\ref{fig:g4iw_file}. The number of non-zero elements in $G^{(2)}$ depends,
in particular, on the type of interactions used.
There is an increasing number of elements from density-density to Kanamori to full Coulomb interaction.

\begin{figure}[]
\centering
\includegraphics[width=9cm]{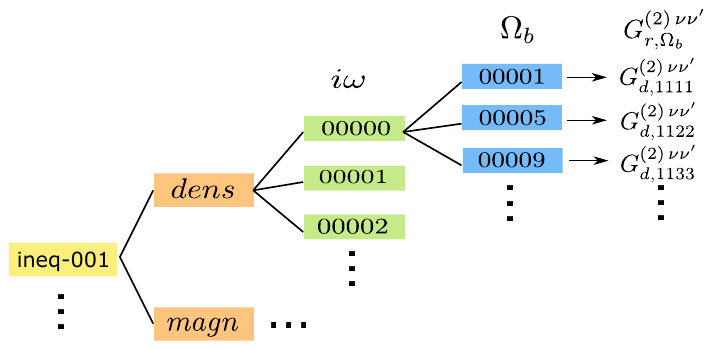}
\caption{Group structure of the {\tt{2PFile}}, which contains the SU(2)- and orbital-symmetrized $G^{(2)}$
in the density and magnetic channel. $i\omega$ refers to the bosonic frequency index,
while $\Omega_b$ is a combined index of the four orbital indices $lmm\pr l\pr$.
Shown are the first few entries in the SrVO$_3$ file (three orbitals, Kanamori interaction).}
\label{fig:g4iw_sym}
\end{figure}

\paragraph*{Group structure of the \tt{2PFile}} The preprocessing program \verb=setupvertex= symmetrizes the two-particle Green's function stored in \verb=G2File= and transforms it into the density and magnetic channel $r\in\{d,m\}$ (see Sec.~\ref{Sec:Setup}).
The symmetrized $G^{(2)}_r$ is then written into the file \verb=2PFile= (see Fig.~\ref{fig:abinitdga_flow_new} and \ref{app:configs}).
The group structure of the latter is shown in Fig.~\ref{fig:g4iw_sym}. The file contains
groups for inequivalent atoms, and subsequent groups for the density and the magnetic channel.
As the AbinitioD$\Gamma$A algorithm is parallelized over the bosonic Matsubara frequency, the data is further split in subgroups for each $\omega$,
allowing for an improved read-in of a given bosonic frequency slice of $G^{(2)}_r$ (see Table \ref{tab:programflow}).
Each bosonic frequency group finally contains subgroups with the non-zero
orbital components of $G^{(2)\, \nu\nu'}_{r, lmm\pr l\pr}$.
These orbital subgroups are labeled by the combined orbital index $\Omega_b$.
The latter is defined through a similar index transformation as in Eq.~\eqref{ind_trafo}, but involving only the four orbital indices, i.e.,
\begin{equation} \label{eq:Omega_b_trafo}
  lmm\pr l\pr \leftrightarrow \Omega_b .
\end{equation}
\noindent
Due to the mapping of six spin components into the two channels,
the size of the \verb=2PFile= is only about one third of the initial \verb=G2File=.

\subsubsection{Compound indices and matrix operations}
\label{Sec:compoundind}

\begin{figure}[]
\centering
\includegraphics[width=14.5cm]{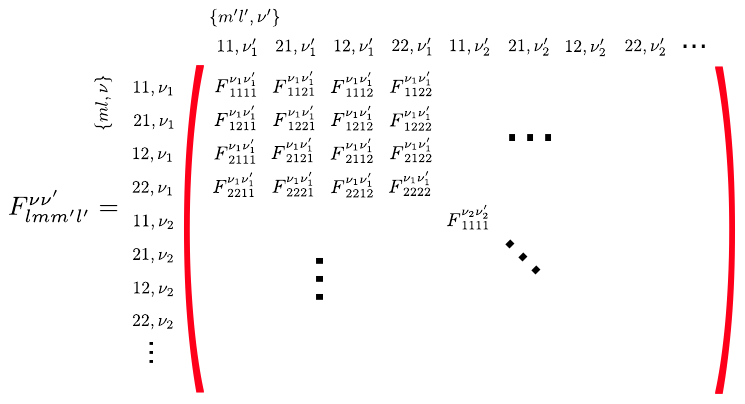}
\caption{By using compound indices $\{ml,\nu \}$ and $\{m\pr l\pr,\nu' \}$,
  $F^{\omega\nu\nu'}_{lmm\pr l\pr}$ can be written in matrix form
  (in the graphics, the "external" bosonic frequency $\omega$ has
  been omitted for simplicity). Explicitly shown is the first orbital block
  ($\nu=\nu_1$, $\nu'=\nu_1'$) for the case of two orbitals.
  Please note that many entries are zero if density-density or Kanamori interactions are employed,
  e.g., $F^{\nu_1\nu_1'}_{1121}=0$ and $F^{\nu_1\nu_1'}_{1112}=0$.}
\label{fig:g_matrix_1}
\end{figure}
\begin{figure}[]
\centering
\includegraphics[width=14.5cm]{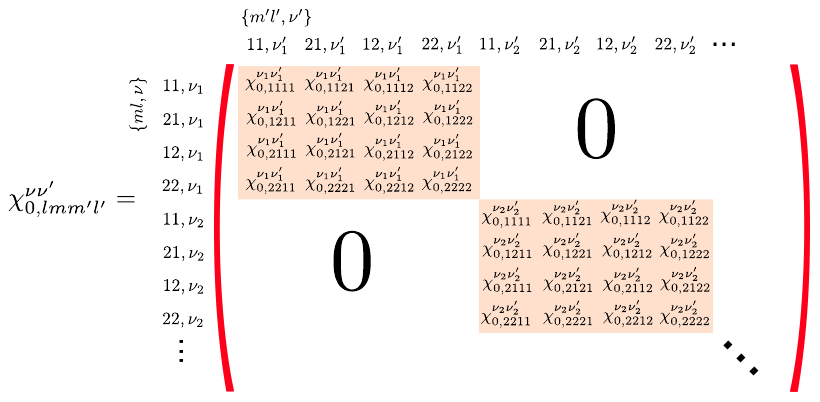}
\caption{Matrix structure of the bubble terms $\chi^{nl,\cvec{q}\nu\nu}_{0,lmm\pr l\pr}$, $\chi^{\cvec{q}\nu\nu}_{0,lmm\pr l\pr}$ and $\chi^{\omega\nu\nu}_{0,lmml}$ (in the graphics, the "external" bosonic index $\cvec{q}=(\kvec{q},\omega)$ has been omitted).
 Shown are the first two "orbital blocks" (in orange). The block-diagonal structure arises from the fact that the bubble terms are diagonal with respect to the fermionic frequency $\delta_{\nu\nu'}$. Here, the full structure is only shown for clarity, the main AbinitioD$\Gamma$A program stores and works only with the non-zero orbital blocks.}
\label{fig:g_matrix}
\end{figure}
\noindent
In order to efficiently perform the orbital and frequency summations, we introduce compound indices
so as to write the equations presented in Section \ref{equations} as matrix operations. The compound indices are obtained by transforming the four orbital and two fermionic frequency indices, e.g., of $F^{\omega\nu\nu'}_{lmm\pr l\pr}$, into two compound indices: the two left orbital indices $lm$ and the left fermionic frequency index $\nu$ are combined into one compound index $\{ml,\nu \}$, while the two right orbital indices $m\pr l\pr$ and the right fermionic frequency index $\nu'$ form the second compound index $\{m\pr l\pr,\nu' \}$.%
\footnote{Note that the bosonic frequency $\omega$ does not enter the compound index; in fact, the AbinitioD$\Gamma$A program is parallelized over the "external" index $\cvec{q} = ( \kvec{q}, \omega )$.}
 This way, $F^{\omega\nu\nu'}_{lmm\pr l\pr}$ can be written in matrix form, $F^{\omega}_{\{ml,\nu'\}\{m\pr l\pr,\nu\}}$, as illustrated in Fig.~\ref{fig:g_matrix_1}. Please note that in the local $F^{\omega\nu\nu'}_{lmm\pr l\pr}$ many matrix elements are zero, since the Kanamori interaction allows only for entries with pairwise matching orbitals. These zero matrix elements are exactly the orbital components not present in \verb=2PFile=. However, in order to perform straightforward matrix operations, one needs to work with the whole matrix including all zero elements. Please note however that, despite this general implementation, due to the spin diagonalization into the density and magnetic channel a SU(2) symmetric interaction is still required.

Similar to the local, full vertex function $F^{\omega\nu'\nu}$, also the bubble terms $\chi^{nl,\cvec{q}\nu\nu}_0$, $\chi^{\cvec{q}\nu\nu}_0$ and $\chi^{\omega\nu\nu}_0$ can be written in matrix form with respect to the compound indices $\{ml,\nu \}$ and $\{m\pr l\pr,\nu' \}$, as visualized in Fig.~\ref{fig:g_matrix}. Since the bubble terms are diagonal with respect to the fermionic frequency indices $\chi_0^{\nu\nu'}=\chi_0^{\nu\nu}\delta_{\nu\nu'}$, they have a block-diagonal structure in the compound basis.

\begin{figure}[]
\centering
\includegraphics[width=12cm]{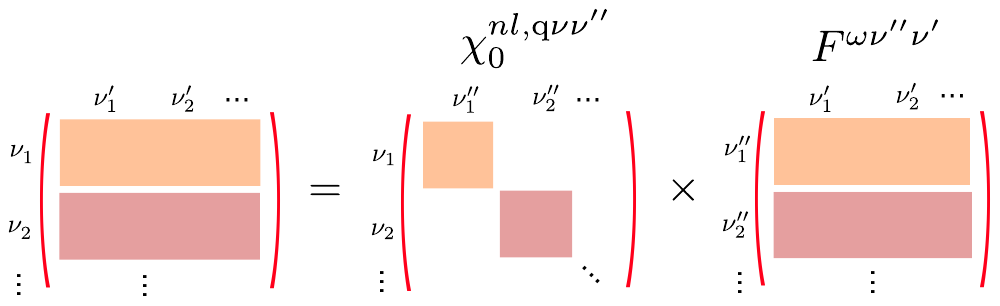}
\caption{The matrix multiplication of the block-diagonal $\chi^{nl,\cvec{q}\nu\nu}_{0,lmm\pr l\pr}$ with the full $F^{\omega\nu\nu'}_{r,lmm\pr l\pr}$ in Eqs.~\eqref{eq:gamma_q_imp} and~\eqref{eqn:eta_imp_mat} can be split into several smaller operations. By multiplying each orbital block of $\chi^{nl,\cvec{q}\nu\nu}_{0,lmm\pr l\pr}$ with the corresponding slice of $F^{\omega\nu\nu'}_{r,lmm\pr l\pr}$ (marked with the same color), multiplications with entries that are zero by construction can be avoided.}
\label{fig:matmul_slice}
\end{figure}

\paragraph*{Simplified matrix operations}
The computation of the three-leg vertices $\gamma^{\cvec{q}}$ and $\eta^{\cvec{q}}$ in Eqs.~\eqref{eq:gamma_q_imp} and~\eqref{eqn:eta_imp_mat} involves a multiplication of $\chi^{nl,\cvec{q}}_0$ with $F^{\omega}$. This matrix multiplication can be simplified by exploiting the block-diagonal structure of $\chi^{nl,\cvec{q}}_0$. In fact, by multiplying each orbital block of $\chi^{nl,\cvec{q}}_0$ with the corresponding horizontal slice of $F^{\omega}$, as shown in Fig.~\ref{fig:matmul_slice}, one can avoid
multiplications involving entries that are zero by construction.

The matrix inversion in the equation for $\eta^{\cvec{q}}$, Eq.~\eqref{eqn:eta_imp_mat}, instead cannot make use of a block-diagonal structure so that the inversion of the full matrix is needed. From a numerical point of view, this matrix inversion is one of the most demanding operations in the main AbinitioD$\Gamma$A program.

\begin{figure}[]
\centering
\includegraphics[width=9cm]{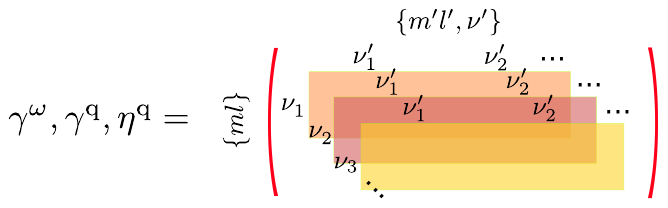}
\caption{Schematic representation of the sum over the left fermionic frequency needed to obtain the three-leg vertices $\gamma^{\omega}$, $\gamma^{\cvec{q}}$ and $\eta^{\cvec{q}}$ in Eqs.~\eqref{eq:gamma_loc_imp}-\eqref{eqn:eta_imp_mat}. By summing over all stacked slices (different colors symbolize different left fermionic frequencies $\nu_i$), the first dimension of the matrix is reduced to the orbital-only compound index $\{lm\}$.}
\label{fig:sum_left_nu}
\end{figure}

The calculation of the three-leg vertices in Eqs.~\eqref{eq:gamma_loc_imp}-\eqref{eqn:eta_imp_mat} furthermore requires a sum over the left fermionic frequency. The summation over this left fermionic frequency makes them, diagrammatically, {\em three-leg} (electron-boson) vertices.
In terms of compound matrices, this sum over the left fermionic frequency is visualized in Fig.~\ref{fig:sum_left_nu}. Through the sum, the left compound index is reduced to an orbital compound index $\{lm\}$ and the resulting matrix is not quadratic any more.
Please note that this summation over the left fermionic frequency needs to be performed explicitly only in order to obtain $\gamma^{\cvec{q}}$ and $\gamma^{\omega}$.\footnote{If the purely local three-leg vertex $\gamma^{\omega}$ is directly computed in CT-HYB, the current sum over the left fermionic frequency to obtain $\gamma^{\omega}$ is redundant.} The three-leg structure of $\eta^{\cvec{q}}$ is actually obtained in a different way, namely by multiplying with $(\vec{\idmatrix} + \gamma^{\omega})$ from the left, as can be seen in Eq.~\eqref{eqn:eta_imp_mat}.

\begin{figure}[]
\centering
\includegraphics[width=10cm]{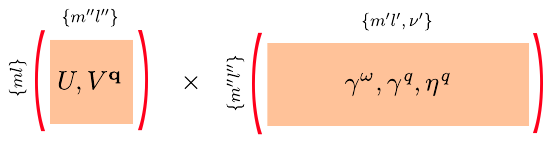}
\caption{Schematic representation of the matrix multiplication between the interaction matrices $U$ and $V^{\kvec{q}}$ and the three-leg vertices $\gamma_r^{\omega}$, $\gamma_r^{\cvec{q}}$ and $\eta_r^{\cvec{q}}$. This operation is part of the Schwinger-Dyson equation of motion~\eqref{eq:eom_final_imp}.}
\label{fig:u_gamma}
\end{figure}

In the equation of motion~\eqref{eq:eom_final_imp}, the three-leg vertices $\gamma_r^{\omega}$, $\gamma_r^{\cvec{q}}$ and $\eta_r^{\cvec{q}}$ are multiplied with the corresponding local and non-local Coulomb interaction terms ($U$, $\tilde{U}$ and $V^{\kvec{q}}$). In order to perform this operation in the basis of compound indices, the four-index $U_{lmm\pr l\pr}$ and $V^{\kvec{q}}_{lmm\pr l\pr}$ are transformed to compound indices $\{ml\}$ and $\{m\pr l\pr\}$. Then, the multiplication of $V^{\kvec{q}}$ and $U$ times the three-leg $\gamma$'s and $\eta$ can be performed easily, as schematically depicted in Fig.~\ref{fig:u_gamma}.

The final convolution with the non-local Green's function $G^{\cvec{k}-\cvec{q}}$ in the equation of motion~\eqref{eq:eom_final_imp} instead is more straightforward to perform by breaking up the compound indices into single orbital and frequency indices.

\subsubsection{Numerical effort}
\label{Sec:numericaleffort}

Often the numerical effort for calculating the local vertex in CT-HYB
is the computationally most demanding task of an AbinitioD$\Gamma$A calculation. The calculation of the vertex scales as $\beta^5(\#o)^4$ with a large prefactor because of the Monte-Carlo sampling (let us remind the reader that $\#o$ is the number of orbitals and $\#\omega\sim\beta$ the number of Matsubara frequencies). This scaling can be understood from the fact that we need to calculate $(\#\omega)^3(\#o)^4$ different components of the local vertex, and the
update of the CT-HYB hybridization matrix requires $\sim \beta^2$ operations because the mean expansion order and hybridization matrix dimension is $\sim \beta$. Since we eventually calculate the self-energy with only one frequency and two orbitals, a higher noise level can be tolerated if $\#\omega$ and $\#o$ are large. Hence, in practice a weaker dependence on $\#\omega$ and $\#o$ is possible.
 Calculating the vertex for SrVO$_3$ with $\#o=3$, $\#\omega=120$ and $\beta=10\,$eV$^{-1}$ took 150000 core h on an Intel Xeon E5-2650v2 (2.6 GHz, 16 cores per node).
One can also employ the asymptotic form~\cite{kaufmann_vertex_asymp_2017,Li_parquet_2016,wentzell_asymptotics_arxiv} of the vertex for large frequencies. This asymptotic part depends on only two frequencies and thus scales as $\beta^4(\#o)^4$. This way the full CT-QMC calculation of the three-frequency vertex can be restricted to a small frequency box, and room temperature calculations are feasible.

Let us now turn to the main AbinitioD$\Gamma$A program itself which is parallelized over the compound index $\cvec{q} = (\kvec{q},\omega)$.
This parallelization gives us a factor $\#q$ $\#\omega$ for the numerical effort ($\#q$: number of $\kvec{q}$-points). For each $\cvec{q}$-point,
 the numerically most costly task is the matrix inversion in Eq.\ (\ref{eqn:eta_imp_mat}). The dimension ${\cal N}$ of the matrix that needs to be inverted is ${\cal N}=\#\omega (\#o)^2$. While simple matrix inversions scale as
${\cal N}^3$ more efficient ones scale roughly as ${\cal N}^{2.5}$.\footnote{The matrix inversion in Eq.~\eqref{eqn:eta_imp_mat} is performed by using the {\tt{lapack}} routines {\tt{zgetrf}} and {\tt{zgetri}}, which compute the inverse of a matrix by triangular decomposition.} Hence, the overall effort is $\sim\#q\#\omega^{3.5}\#o^5$.
The numerical effort for calculating the self-energy via the equation of motion (\ref{eq:eom_final_imp}), on the other hand,
 is $\sim\#q^2\#\omega^2\#o^6$ and only
becomes the leading contribution at high temperatures and for a large number of q-points.

\begin{figure}[]
\centering
\includegraphics[width=7cm]{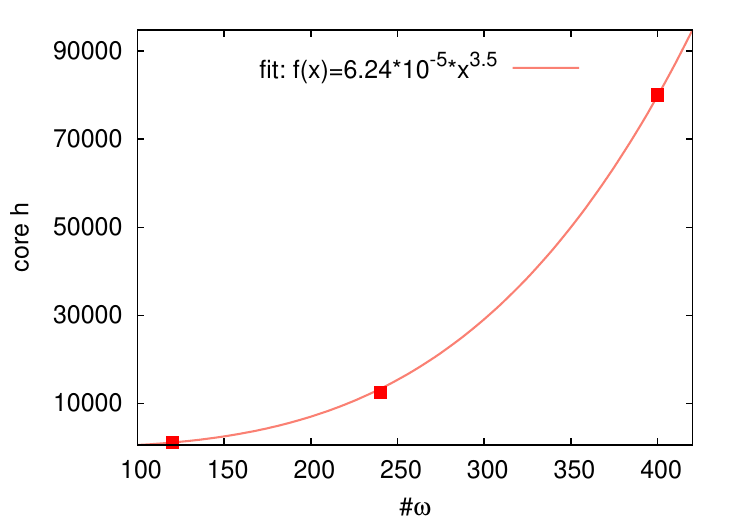}
\caption{Computational effort of the AbinitioD$\Gamma$A program with respect to the number of involved Matsubara frequencies $\#\omega$.}
\label{fig:cpu}
\end{figure}

For the AbinitioD$\Gamma$A computation of SrVO$_3$ with $\#o=3$ and $\#q=20^3$, the numerical effort with respect to the number of Matsubara frequencies $\#\omega$ has explicitly been tested by performing computations with three different frequency box sizes: $\#\omega=120$, $\#\omega=240$ and $\#\omega=400$. Fig.~\ref{fig:cpu} shows the respective numerical effort in core h. From Fig.~\ref{fig:cpu} it can be seen that the main AbinitioD$\Gamma$A program indeed roughly scales with $(\#\omega)^{3.5}$.

\section{Conclusion and outlook}
\label{conc}
\label{Sec:Conclusion}
\noindent
In this paper we have outlined the structure of the
AbinitioD$\Gamma$A program package and
provided information on how to use it. While the numerical effort
is considerably larger than
for state-of-the-art DFT+DMFT calculations,
our code makes realistic multi-orbital post-DMFT
studies feasible. We expect AbinitioD$\Gamma$A
calculations to provide valuable insight into the physics of non-local correlations in
strongly correlated materials.
Besides studying, e.g., the nature of non-local spin-fluctuations, our methodology
also allows to systematically assess the error made in DFT+DMFT calculations.

\paragraph*{Approximations}
Let us, at this point, reflect on the approximations involved in AbinitioD$\Gamma$A. As a matter of course, any method aiming at an {\em ab initio} calculation of materials, requires approximations.
First of all, there is the D$\Gamma$A approximation which assumes the irreducible two-particle vertex to be local. In a more complete variant, this is the local two-particle fully irreducible vertex $\Lambda$ and the parquet equations are employed to calculate the non-local full vertex and self-energy in parquet D$\Gamma$A \cite{Li2016,Li2019}. This approach is numerically too involved for realistic multi-orbital calculations. Instead AbinitioD$\Gamma$A uses the Bethe-Salpeter ladder diagrams in both, the particle hole and transversal particle-hole channel, starting with the local irreducible vertex $\Gamma$ in the respective channel (eventually the Bethe-Salpeter equations are reformulated in terms of the local full vertex $F$). That is, the particle-particle channel and with it superconducting fluctuations and weak localization effects are neglected, as is the feedback between the two particle-hole channels beyond the local level.

Second, while AbinitioD$\Gamma$A includes all the DMFT diagrams, all the $GW$ diagrams if the non-local Coulomb interaction is included in the AbinitioD$\Gamma$A starting vertex $\Gamma$, as well as non-local spin fluctuations and further diagrams, we still need to supplement AbinitioD$\Gamma$A with $GW$ or DFT outside a low energy window of $\sim 10$ orbitals. This bears the danger of a possible overscreening of the Coulomb interaction in the constrained random phase approximation as calculations for simple models suggest \cite{Honerkamp2018}.
 Further, so far only a non-local $V^q$ is foreseen, not the $k$- and $k'$-dependence thereof (which would result in considerably more involved Bethe-Salpeter equations).

Third, if non-local correlations become truly large, actually two self-consistencies are needed: (i) the self-consistent calculation of the Green function lines connecting the vertex blocks and (ii) the self-consistent recalculation of the irreducible vertex. We plan these self-consistencies as additional scripts outside the core AbinitioD$\Gamma$A code introduced in the present paper, in future versions of the code. A prospective alternative to mimic these self-consistencies is the so-called $\lambda$ correction \cite{Toschi2007,Katanin2009} or a ${\cal U}_{\omega}$ \cite{Krien2019} correction, but for multi-orbital calculations this becomes cumbersome since many $\lambda$ or ${\cal U}_{\omega}$ parameters need to be adjusted.

\paragraph*{Outlook}
Materials calculations with diagrammatic extensions of DMFT
are just at the beginning.
We believe that our code will contribute turning this route into
a thriving research field, similar to what DFT+DMFT is today.
We hope to foster this development by releasing the
AbinitioD$\Gamma$A program package
under the terms of the GNU General Public License version 3.

\section{Acknowledgments}
\noindent
We are deeply indebted to Patrik Gunacker for furthering the w2dynamics code to calculate multi-orbital vertices,
for fruitful discussions and the cooperation within the SrVO$_3$ project.
This work has been financially supported by the European Research
Council under the European Union's Seventh Framework Program
(FP/2007-2013) through ERC grant agreement n.\ 306447. Calculations have
 been done on the Vienna Scientific Cluster~(VSC).





\bibliographystyle{elsarticle-num}
\bibliography{main,Bibliography_CPC}

\newpage\clearpage
\appendix







\newpage
\clearpage
\section{Exemplary config files}
\label{app:configs}
\subsection{Config file for a self-energy calculation}
\label{app:config1}
\noindent
In the following we show the config file, as used in the main text. Here we calculate both the
self-energies \\
(\verb_calc-eom = T_) and the susceptibilities (\verb_calc-susc = T_) on the largest
possible momentum grid and the frequency box. The interaction is defined by providing Kanamori parameters (\verb_Udd_, \verb_Vdd_ and \verb_Jdd_).
\\[\baselineskip]\noindent
\label{app:configself}
\begin{verbatim}
[General]
calc-susc = T # calculate the momentum-dependent susceptibilities
calc-eom  = T # calculate the dga-selfenergy via the equation of motion

# number of positive f/b frequencies used from the vertex
N4iwf = -1 # full box
N4iwb = -1 # full box

NAt = 1 # Number of atoms

HkFile = srvo3_k20.hk # Wannier Hamiltonian

k-grid = 20 20 20 # Wannier Hamiltonian and eom momentum grid
q-grid = 20 20 20 # Grid for susc, and q-sum in eom

[Atoms]
[[1]]
Interaction = Kanamori # interaction type
Nd = 3 # number of d-bands
Np = 0 # number of p-bands
Udd = 5.0 # intra-orbital interaction
Vdd = 3.5 # inter-orbital interaction
Jdd = 0.75 # Hund's coupling

[One-Particle]
1PFile = srvo3-1pg.hdf5 # DMFT 1PG
orb-sym = T

[Two-Particle]
2PFile = srvo3-2pg-symmetrized.hdf5 # symmetrized vertex
vertex-type = 0
\end{verbatim}

\newpage\clearpage
\noindent
\subsection{Config file for a susceptibility calculation}
\label{app:config2}
\noindent
If one is only interested in susceptibilities,
there are a few ways to save computational time. Since in this
case it is not necessary to compute the equation of motion (which requires a momentum sum and a bosonic frequency sum (see Eq.~\eqref{eq:eom_final_imp}), the user
has the freedom to reduce the number of bosonic frequencies $\omega$
and momenta $\kvec{q}$.\\
In the following we show an example config file that contains
the settings for calculating static susceptibilities on certain user-specified
$\kvec{q}$-points.
\\[\baselineskip]\noindent
\label{app:configsusc}
\begin{verbatim}
[General]
calc-susc = T # calculate the momentum-dependent susceptibilities
calc-eom  = F # calculate the dga-selfenergy via the equation of motion

# number of positive f/b frequencies used from the vertex
N4iwf = -1 # full fermionic box
N4iwb =  0 # only iwn = 0

NAt = 1 # Number of atoms

HkFile = srvo3_k20.hk # Wannier Hamiltonian

k-grid = 20 20 20 # Wannier Hamiltonian momentum grid
QDataFile = qpath

[Atoms]
[[1]]
Interaction = Kanamori # interaction type
Nd = 3 # number of d-bands
Np = 0 # number of p-bands
Udd = 5.0 # intra-orbital interaction
Vdd = 3.5 # inter-orbital interaction
Jdd = 0.75 # Hund's coupling

[One-Particle]
1PFile = srvo3-1pg.hdf5 # DMFT 1PG
orb-sym = T

[Two-Particle]
2PFile = srvo3-2pg-symmetrized.hdf5 # symmetrized vertex
vertex-type = 0
\end{verbatim}

\noindent
The file name provided in the \verb=QDataFile= variable corresponds to a simple plain text file.
The Bethe-Salpeter equations are then only solved for the given $\kvec{q}$-point(s) instead of a
regular sized grid.
Every line in the given file specifies a $\kvec{q}$-point
by 3 integer numbers (please note that 0 corresponds to the first $\kvec{q}$-point).
The \verb_qpath_ file may have the following contents:
\begin{verbatim}
0  0  0
0  10 0
10 10 0
\end{verbatim}
Considering our example Hamiltonian, \verb=0 0 0=, \verb= 0 10 0= and \verb=10 10 0= refers to the $\Gamma$-,
$X$-, and $M$-point, respectively. Please note that if there is no \verb=QDataFile= provided, the given $\kvec{q}$-mesh provided in \verb=q-grid= is used instead.

\section{Results with the provided data}
\label{app:results}
\begin{figure}[h]
\begin{center}
\includegraphics[scale=0.52]{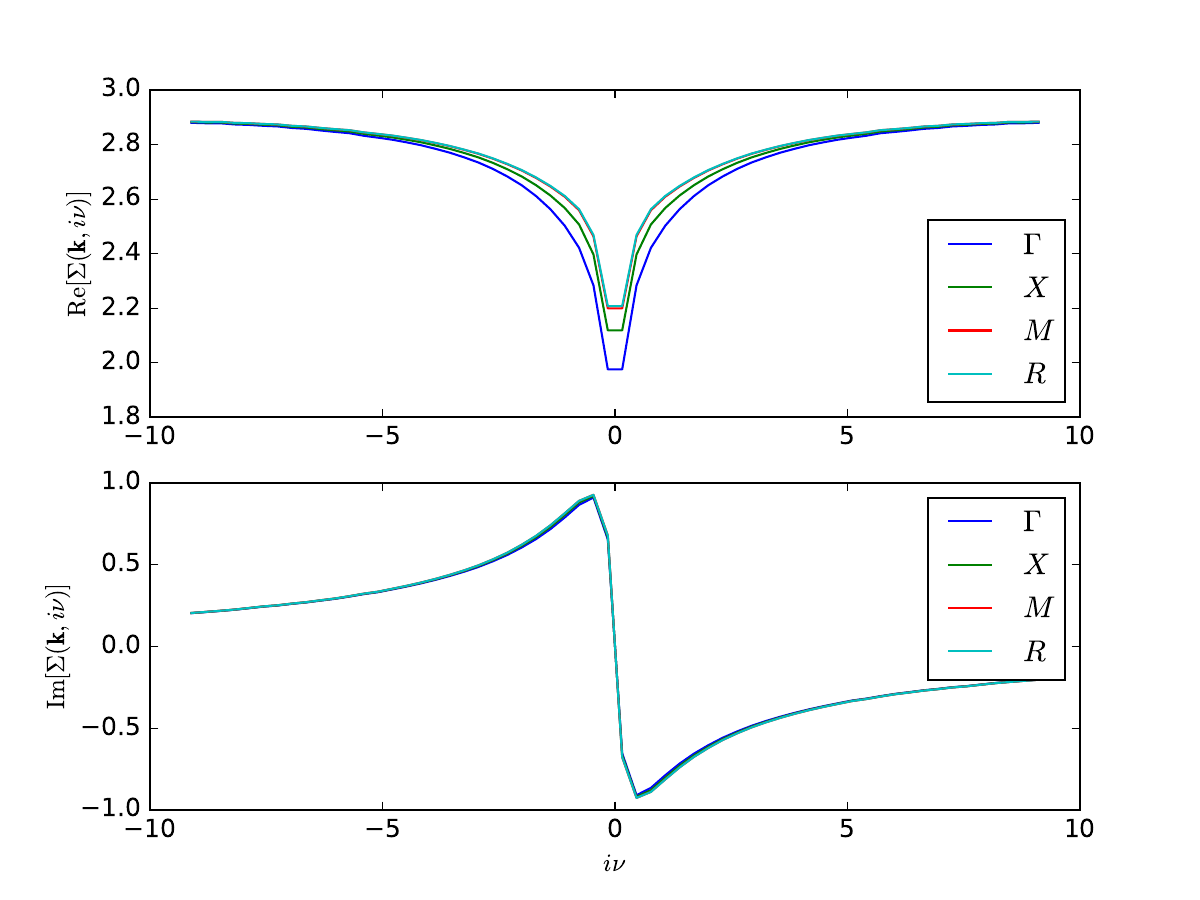}
\caption{Self energy for a q-qrid calculation evaluated at the high symmetry points $\Gamma$, $X$, $M$ and $R$. The plot script can be found in documentation/scripts/1D\_plot\_selfenergy.py}
\label{Fig:Sec:Sigma1}
\end{center}
\end{figure}

\begin{figure}[h]
\begin{center}
\includegraphics[clip, trim=0cm 3cm 0cm 3cm, scale=0.52]{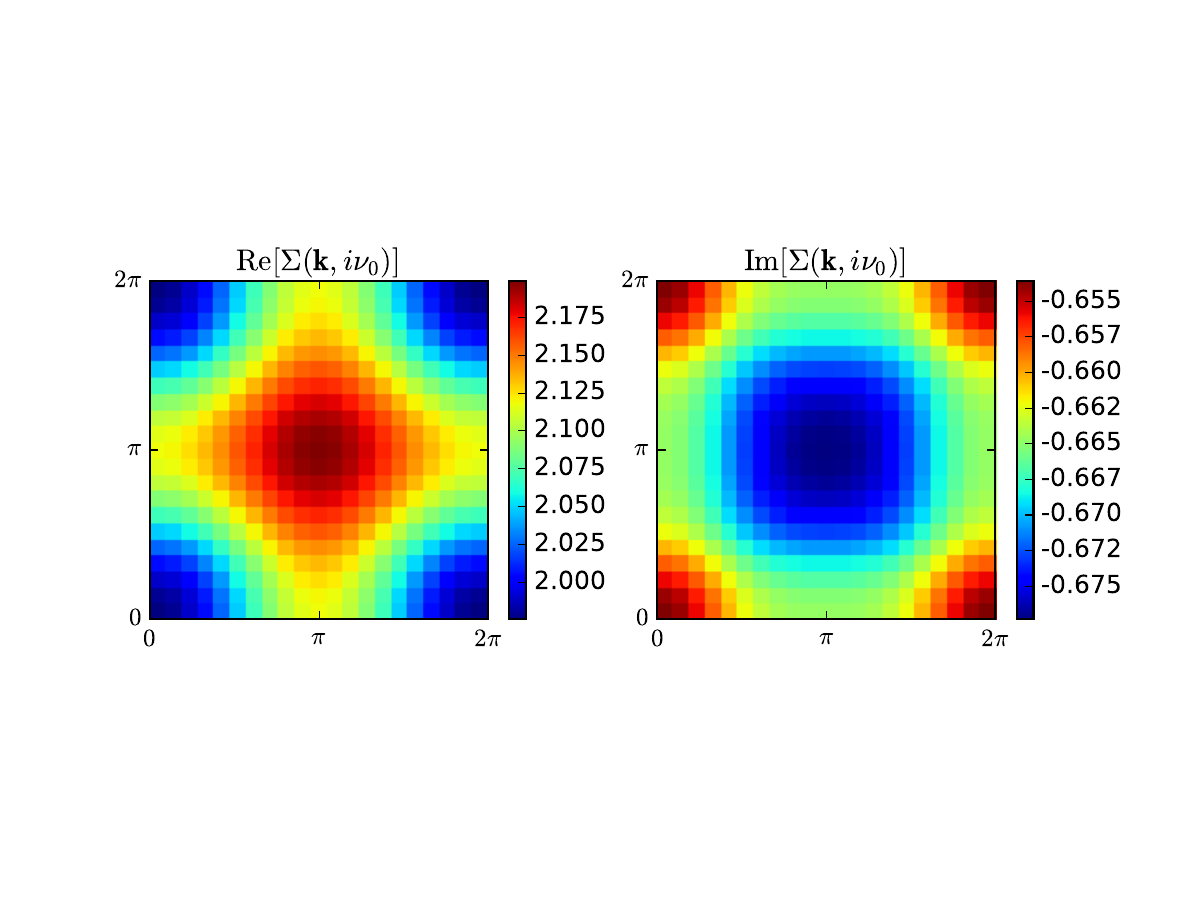}
\caption{Self-energy for a q-qrid calculation in the $k_z=0$ plane where we show the $\mathbf{k}$-resolved value on the first Matsubara frequency $i\nu_0 = \frac{\pi}{\beta}$. Please note that here the $\Gamma$-point is located at the bottom left corner. The plot script can be found in documentation/scripts/2D\_plot\_selfenergy\_plane.py}
\label{Fig:Sec:Sigma2}
\end{center}
\end{figure}

\noindent
Here we present the results for the test data from the repository (\verb+srvo3-testdata/+). The data shown in Sec.~\ref{Sec:example}
were calculated on a $\mathbf{k}$-grid of $20x20x20$ with a frequency box of 200 positive fermionic and bosonic frequencies while
the test data provided in the repository only has a frequency box of 30 positive fermionic and bosonic frequency.
Using the same config file (see \ref{app:config1}), as well as the plot scripts
found in \verb+documentation/scripts/+, we obtain the self-energy displayed in Fig.~\ref{Fig:Sec:Sigma1} and \ref{Fig:Sec:Sigma2}. Note, that these results are not yet converged with respect to the size of the frequency box.

\begin{figure}[h]
\begin{center}
\includegraphics[scale=0.50]{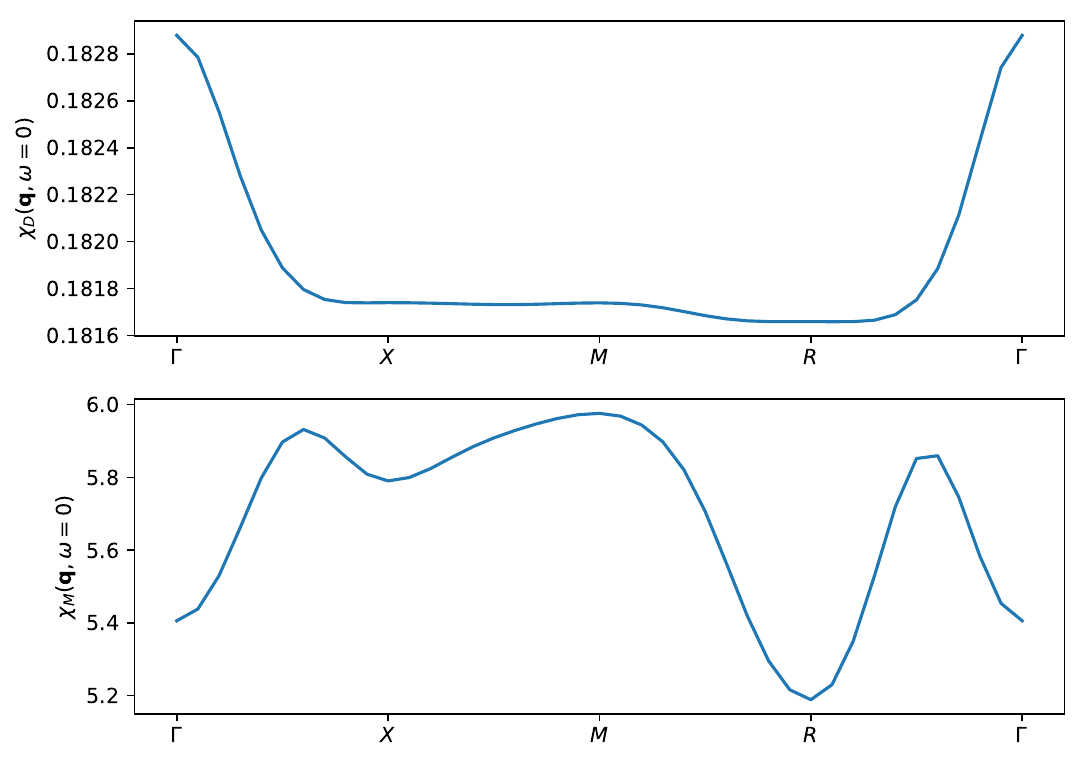}
\caption{Susceptibility for a q-path calculation. The plot script can be found in documentation/scripts/1D\_plot\_qpath.py.
Please note that the generally smaller susceptibilities (compared to Sec.~\ref{Sec:example}) are caused by the much smaller box size.}
\label{Fig:Sec:chi}
\end{center}
\end{figure}

\newpage
\noindent
By using the second provided config file (see \ref{app:config2}) we obtain the susceptibilities of Fig.~\ref{Fig:Sec:chi}.
Please keep in mind that the purpose of this test calculation is to provide a fast, computationally inexpensive check of the AbinitioD$\Gamma$A code;
30 (positive) Matsubara frequencies are not enough to arrive at a result that is converged with respect to the frequency box.
The effect of an insufficient number of Matsubara frequencies is most prominent in the frequency-summed susceptibility.

\end{document}